\documentclass[twocolumn]{aastex631}
\usepackage{lmodern}

\def\Reflex{R{\sc efle}X}
\def\feka{$\rm Fe\,K\alpha$}

\def\h2{$H_2$}

\def\rxtopo{\textsc{RXToPo}}
\def\rxagn{\textsc{RXagn1}}
\def\rxtorus{\textsc{RXTorus}}

\usepackage{multirow}
\usepackage{threeparttable}
\usepackage{subfigure}

\begin{document}

\title{The study of the circumnuclear environment of accreting supermassive black holes with realistic X-ray spectral models}

\author[0009-0002-4945-5121]{G. Dimopoulos}
\affiliation
{Instituto de Estudios Astrof\'{\i}sicos, Facultad de Ingeniería y Ciencias, Universidad Diego Portales, Avenida Ejercito Libertador 441, Santiago, Chile}
\author[0000-0001-5231-2645]{C. Ricci}
\affiliation
{Instituto de Estudios Astrof\'{\i}sicos, Facultad de Ingeniería y Ciencias, Universidad Diego Portales, Avenida Ejercito Libertador 441, Santiago, Chile}
\affiliation{
Kavli Institute for Astronomy and Astrophysics, Peking University, Beijing 100871, China}
\author[0000-0002-8108-9179]{S. Paltani}
\affiliation{
Department of Astronomy, University of Geneva, ch. d’Ecogia 16, 1290, Versoix, Switzerland}



\begin{abstract}

X-ray spectral modeling is a powerful tool for studying the immediate environment of accreting objects, including supermassive black holes.
Several models, either phenomenological or physically driven, have been developed over the past decade to study X-ray spectra, delivering important insights into the properties of circumnuclear material of active galactic nuclei (AGN).
Despite the fact that these models are able to reproduce the data well, they often lack realistic geometries, and most of them consist of simplified configurations such as a slab or a torus.
We use the ray-tracing code \textsc{RefleX} to generate new spectral models that cover a wide energy range in the X-ray band, adopting a realistic configuration for the surrounding material.
We introduce two new table models that are publicly available:
1) the RXToPo model, which features an X-ray source along with a dusty torus and a polar hollow cone;
2) the RXagn1 model, which includes, besides the torus and polar cone, also the accretion disk and the broad line region.
Both models were applied to the X-ray spectrum of NGC 424, demonstrating their potential to study sources whose X-ray emission is dominated by reprocessed radiation.

\end{abstract}



\section{Introduction}\label{sec:intro}

Supermassive black holes (SMBHs) are known to reside at the center of most galaxies (e.g., \citealt{Salpeter_earlySMBH:1964ApJ,Lynden-Bell_earlyQSO:1969Natur,CavalierePadovani_AGNvsNormal:1989ApJ,Marconi_BH_demographics:2004,Kormendy_Review_SMBH:2013}).
These SMBHs are often embedded in large amounts of gas and dust (e.g., \citealt{RamosAlmeida&Ricci:2017Nat}), which serve as the necessary reservoir of material that is eventually funneled onto the SMBH during the accreting phase, when these objects are observed as active galactic nuclei (AGN).
The accretion disk that is formed during this phase can produce a significant fraction of the photons that come out of these systems (e.g., \citealt{NovikovThorne:1973,Shakura&Sunyaev:1973}).
The UV/optical photons produced in the accretion disk are then up-scattered to the X-rays through inverse Compton scattering by a hot electron cloud located very close to the SMBH (e.g., \citealt{ShapiroX-rays:1976ApJ, Sunyaev_Titarchuk:1980,Lightman_Zdiarski:1987ApJComptonScattering,Haardt&Maraschi:1991,Haardt&mar:1993,Fabian:2009}).
This hot electron cloud, known as the X-ray corona, produces most of the X-ray radiation in these systems.

Although we do not yet have a complete understanding of the immediate environments of SMBHs, a widely accepted theory suggests that a thick torus of dust and gas obscures the central region. This paradigm, known as the unification model, suggests that different types of AGN arise when viewing this structure at different angles
(e.g., \citealt{Antonucci:1993,Urry&Padovani:1995,Netzer_UniModelreview:2015}).
In most AGN ($\sim 60-70\%$) in the local Universe the X-ray source is obscured \citep{Ricci+:2015}. The dusty torus is expected to absorb a significant amount of the disk emission and re-emit it at infrared wavelengths (e.g., \citealt{Rowan-RobinsonIRmodels:1995MNRAS,Rowan-Robinson:1998ASPC,stalevski_dust_2016}).
The accretion disk and the dusty torus are expected to be connected through the broad line region (BLR), where broad emission lines are produced by the high velocity of the gas ($> 2000\,{\rm km\,s^{-1}}$; e.g., \citeauthor{PetersonBLR:1991ApJ} \citeyear{PetersonBLR:1991ApJ,PetersonReverberation:1993PASP,Peterson+BLR:2004ApJ}, \citealt{Kaspi+Reverberation:2000ApJ,Goad+BLR:2012,GravityColab:2020,GRAVITY_BLR:2024AA}).

Torus models have been shown to successfully reproduce the IR emission of local AGN, which further supports the unification model paradigm (e.g., \citealt{Alonso-HerreroIRgalaxies:2009AA,RamosAlmeida_IRSeyferts:2009ApJ,Hao_IR-torus:2005,MendozaCastrejon:IRtorus:2015,Hatziminaoglou+:MIRtorus:2015ApJ}).
In very bright and nearby AGN, such as the Circinus galaxy and NGC\,1068, mid-infrared observations have revealed the presence of dust extending in the polar direction (e.g., \citealt{Honig:PolarCone:2013ApJ,LopezGonzaga:NGC1068:2014,asmus_PolarCOne:2016,Stalevski_PolarCone:2017}). This extended emission could be associated with winds originating from the central region of the AGN, driven by radiation pressure on dust (e.g., \citealp{Fabian:2006,Hoenig:2007,Venanzi:2020}), in agreement with X-ray observations \citep{Ricci:2017Nat, Ricci:2022,Laloux:2024}.
The polar region could contribute significantly to the overall infrared emission from these systems, and is expected to reside outside the dust sublimation region, where dust is thought to exist as a mix of carbon and olivine silicate grains.

When X-ray photons intercept the aforementioned configurations of gas and dust, some of them are scattered, absorbed and/or produce fluorescence (e.g., \citealt{George_Fabian_reflection:1991MNRAS,Magdziarz_Zdiarski_reflection:1995MNRAS,Lamer_Uttley_RXTEreflection:2000MNRAS,Alexander_2013CXB+Reflection:2013,DelMoroNustarSurvey:2017ApJ,Gupta+:2021}). 
One of the most important features of reprocessed X-ray radiation are fluorescent lines, such as the iron K$\alpha$ line (\feka{}) at $\rm 6.4\,keV$ (e.g., \citealp{Nandra_GeorgeASCA_FEKA+:1997ApJ,Yaqoob_Serlemitos_FEKA3C273_2000ApJ,Shu_FEKA_2010,Shu:2011}). Another important spectral feature is the "Compton hump", a broad component typically found at energies around $\rm 30\,keV$ (e.g., \citealt{Fabian_ComptonHump:1991MNRAS}).
Reprocessed X-ray radiation can therefore provide valuable information about the properties of the material surrounding the SMBH, including its composition and geometry (e.g., \citealt{Reynolds_reflection_spin:2008ApJ,BrightmanNandraXMMsurvey:2011MNRAS,Goad+BLR:2012,Garcia+:2013,Saha+:Torus_Xrays:2021,Diaz_reflection_LLAGN:2023AA,Dong+Refelction_Disk:2023}).

To constrain the main physical properties of the circumnuclear material, numerous X-ray models have been developed over the past decades. These models incorporate a variety of geometries and physical processes.
One of the most widely used is \textsc{pexrav} \citep{Magdziarz&Zdziarski:1995}, which considers a corona on top of a slab, representing the accretion disk.
More sophisticated disk-reflection models, such as \textsc{xillver} (\citeauthor{Garcia+:2013} \citeyear{Garcia+:2010}, \citeyear{Garcia+:2013}) and \textsc{rellxil} \citep{Garcia+:2014},
implemented relativistic effects in the accretion disk.

A significant amount of photons that are emitted from the central region are expected to interact with the gas and dust found beyond the accretion disk, and in particular with the dusty torus.
As a result, several torus models have been developed in the past 15\,years.
\textsc{MYtorus} \citep{Murphy&Yaqoob:2009} assumes an axisymmetric toroidal geometry, implementing the classic orientation-dependent AGN paradigm, and considering a wide range of torus equatorial column densities ($N_{\rm H\,eq} = 10^{22}-10^{25}\,{\rm cm^{-2}}$).
The RXTorus model was built using the radiative transfer code \Reflex{} \citep{Paltani&Ricci:2017}.
Similar to the \textsc{MYtorus} it features a toroidal medium, although in this model, besides the equatorial hydrogen column density ($N_{\rm H\,eq}=10^{22}-10^{25}\,{\rm cm^{-2}}$), the covering factor (i.e. the fraction of the sky of the AGN covered by gas and dust) is a free parameter.
The \textsc{borus02} model \citep{Balokovic+:2018} considers a spherical torus model, and it only includes reprocessed X-ray radiation.
The two main free parameters of this model are the torus column density, in the $\log(N_{\rm H}/{\rm cm^{-2}})=22.0-25.5$ range, and the covering factor of the torus. Over the past decade, several models featuring clumpy torus geometries have been developed. \citet{Liu_Li_torusModel:2014ApJ} used the toolkit \textsc{Geant4} \citep{Geant4AGOSTINELLI:2003} to create a model in which the torus is clumpy.
The number of clouds ($N_{\rm clouds}$), the filling factor ($\phi$) and the total line-of-sight column density ($N_{\rm H}$) are free parameters in this model.
\citet{Tanimoto:2019} introduced the XCLUMPY torus model, an X-ray adaptation of the CLUMPY infrared model \citep{Nenkova+:2008}.
The UxClumpy model \citep{XARS_Buchner:2019A&A} is also based on the CLUMPY infrared model, and assumes the same toroidal geometry.
Besides the X-ray spectral models that are available in the literature with their given configuration, there are different Monte-Carlo codes that propagate photons into various media.
Some examples of such codes are: the \texttt{MONACO} (MONte Carlo simulation for Astrophysics and COsmology; \citealp{MONACO:2011}), the python code \texttt{XARS} (X-ray Monte-Carlo simulator; \citealp{XARS_Buchner:2019A&A}), and \texttt{SKIRT}, which spans a large range of energies, from radio to X-rays \citep{skirt_I:2003,skirt_II:2011ApJS,SKIRT_Xrays:2023}.

The aforementioned models are widely utilized in the literature. However, to accurately reproduce the reprocessed X-ray radiation observed in AGN, it is essential to consider all the different structures present in AGN, such as the accretion disk, broad line region, torus, and polar dust. X-ray photons are expected to interact with all these components; thus, simplified approaches, such as separate modeling of the torus and the disk, could suffer from strong degeneracies.


In this work, we use the ray-tracing code \Reflex{} to develop two new complex models to realistically reproduce the circumnuclear material around SMBHs. The first model, \rxtopo{}, includes a polar hollow cone (representing the polar dusty gas) and a torus. The second model, \rxagn{} is an extension of \rxtopo{} and includes, in addition to the torus and polar dust, an accretion disk and a broad line region.
In \S\,\ref{sec:sim_setup} we present \Reflex{}, the code we used to generate our geometrical objects, and briefly summarize its features. In \S\,\ref{sec:rxtopo} and \S\,\ref{sec:rxagn} we introduce the two new spectral models, including all the details necessary for the reader to understand how they were built, as well as their physical motivation.
Next in \S\,\ref{sec:comparison} we present the format of the models along with the different available variations of each model in terms of the physical processes recorded in order to generate them.
We demonstrate the capabilities of these two new models by fitting the combined {\it XMM-Newton} and \textit{NuSTAR} spectra of NGC\,424, a heavily obscured AGN in which the X-ray emission is dominated by reflection features. Finally, we present our summary and conclusions in \S\,\ref{sec:summary}.

\section{Simulation setup}\label{sec:sim_setup}

\subsection{\Reflex{}}\label{sec:reflex}

In order to construct a physically-motivated AGN model, we conducted X-ray spectral simulations using the \Reflex{}\footnote{\url{https://www.astro.unige.ch/reflex/}} ray-tracing code \citep{Paltani&Ricci:2017}.
\Reflex{} models the propagation of X-ray photons with energies ranging from $\rm 0.1\,keV$ to $\sim \rm 1\,MeV$, taking into account all the main physical processes that occur at these energies.
It considers the full relativistic Klein-Nishina formula for the cross-sections as well as the proper corrections when the scattering involves bound atoms.
Moreover, it includes the absorption energy shells for 30 neutral elements along with their respective emission lines.
Finally, the latest version implements dust grains, where the interaction is governed by the Mie theory \citep{Mie_theory:1908AnP,Wiscombe_Mietheory:1980ApOpt}.
The modular design of \Reflex{} allows the user to create X-ray spectral models by incorporating multiple building blocks, which can vary in both their geometrical form and their physical properties.
Through Monte Carlo simulations, \Reflex{} tracks each photon as it travels through the material, marking all interactions that occur along the way.
The code generates images and spectra based on the chosen geometrical setup; however, in this work, we focus only on the spectral output to develop our models. In \S\,\ref{sec:rxtopo} and \S\,\ref{sec:rxagn} the geometries used are described in detail.
We utilized \Reflex{}\,3.0 \citep{Ricci&Paltani:2023}, which also includes interactions between photons and dust grains.

\subsection{The RXToPo model}\label{sec:rxtopo}

\citet{Paltani&Ricci:2017} along with the release of the \Reflex{} code, published the table model \textsc{RXTorus}, which implements a toroidal reprocessing medium and the X-ray source.
The free parameters are the torus column density and covering factor, as well as the photon index of the primary X-ray continuum.
The covering factor is defined as the ratio \textit{r/R} between the inner cutout radius of the doughnut \textit{r} and the distance from the center of the simulation \textit{R}, i.e. the SMBH.
The torus is expected to be composed of a mixture of gas and dust, and to reside outside the sublimation zone (e.g., \citealt{Mor:2009,GravityNGC1068:2020}).
Therefore, more recently an updated version of this toroidal model, including scattering and absorption associated with dust grains, \textsc{RXTorusD} \citep{Ricci&Paltani:2023}, was released.

\begin{figure}[ht!]
    \centering
    \begin{subfigure}
        \centering
        \includegraphics[width=0.9\columnwidth]{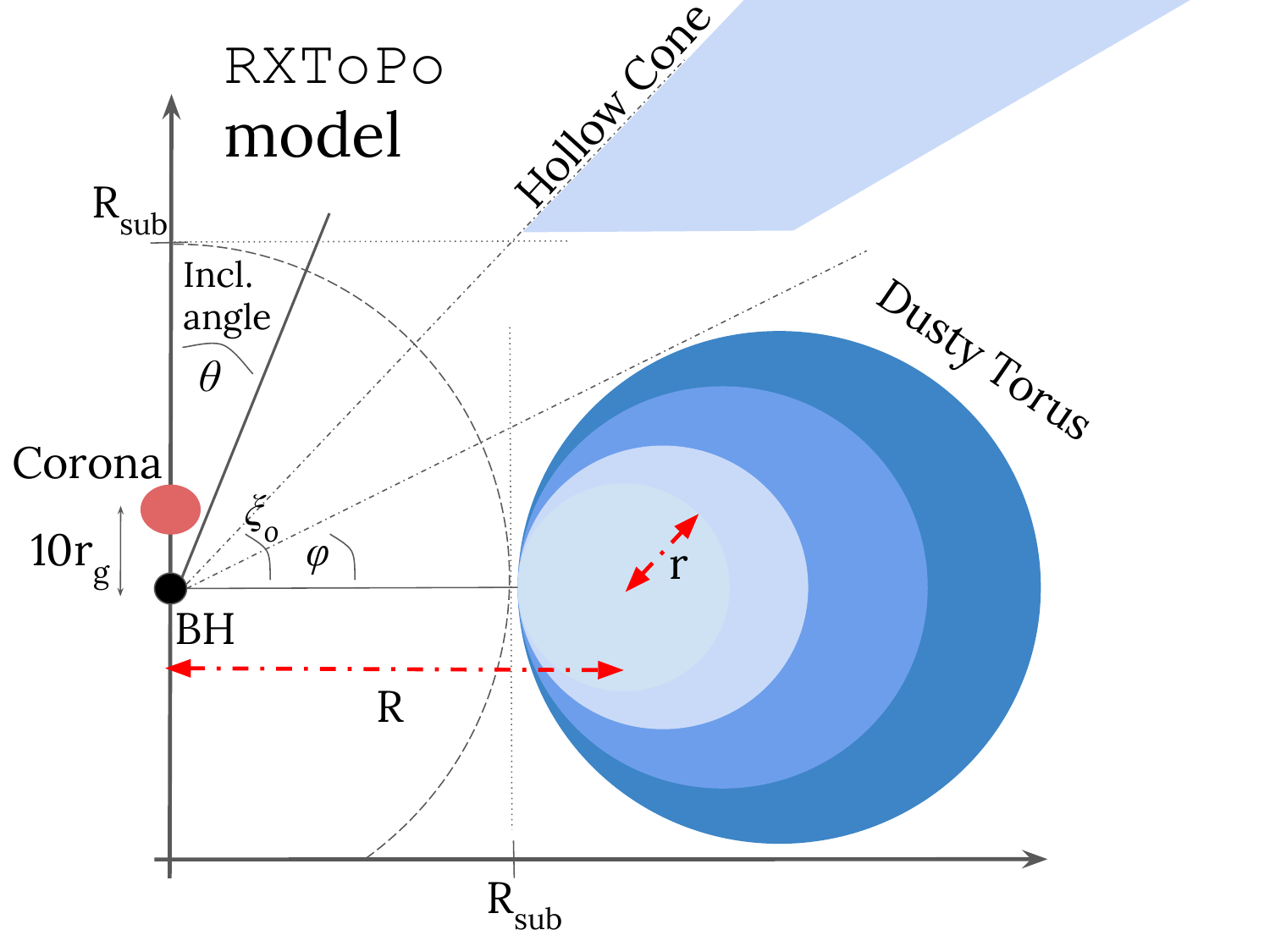}
    \end{subfigure}
    \hfill
    \begin{subfigure}
        \centering
        \includegraphics[width=0.9\columnwidth]{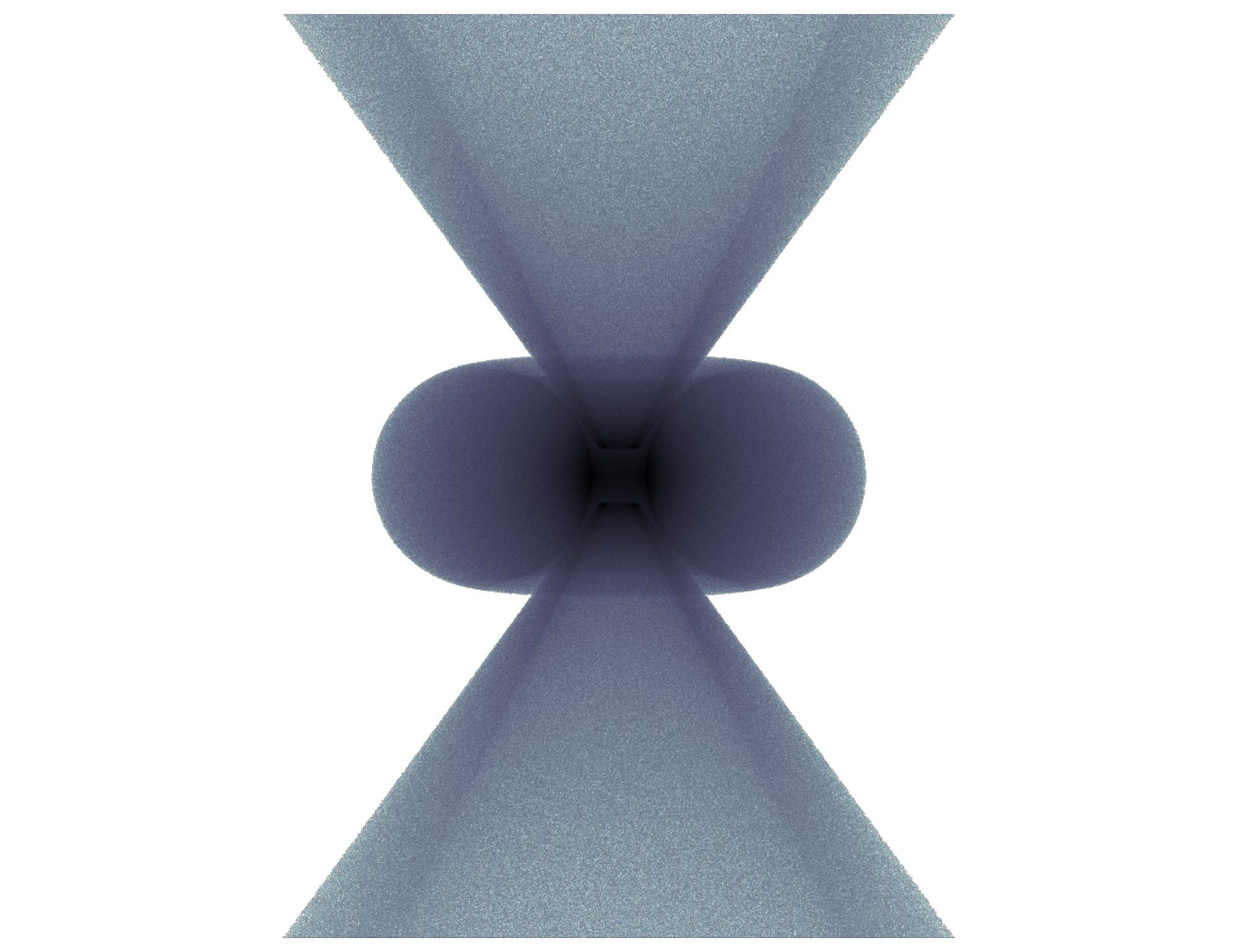}
    \end{subfigure}
    \caption{(Top) A visual illustration of the configuration of the \rxtopo{} model. In the center of the system is the SMBH with the corona placed on top of it. The torus can be found after the sublimation zone with a variable size. Perpendicular to the plane of the torus is the hollow polar cone. The dash-dotted line (red) depicts the R and r that are used to calculate the covering factor of the torus. The components are not scaled. The angle $\theta$ is the observing angle of the system. $\phi$ is the opening angle of the torus and $\xi_{\circ}$ is the maximum angle of the polar hollow cone. (Bottom) A cutout generated using \Reflex{} showing a torus with $CF^{\rm tor}=0.6$ and the polar hollow cone with $10^{\circ}$ angular width.}
    \label{fig:rxtopo_cartoon}
\end{figure}

We introduce here a new model, \rxtopo{}, which can be considered as an extension of \textsc{RXTorusD}.
This model includes, in addition to the dusty torus, a polar dusty cone perpendicular to the accretion disk.
In Fig.\,\ref{fig:rxtopo_cartoon} a cartoon of the model is shown, featuring all the aforementioned components.
The \rxtopo{} model has five free parameters (see Table\,\ref{Tab:rxtopo components}): i) the photon index ($\Gamma$) of the primary power-law continuum, which varies from 1.6 to 2.4; ii) the observing angle of the system ($\theta_{\circ}$), which extends from zero degrees (face-on view) to 90 degrees (edge-on view); iii) the equatorial column density ($N_{\rm H}^{\rm tor}$) and iv) the covering factor ($CF^{\rm tor}$) of the dusty torus (details in \S\,\ref{sec:torus}); v) the column density of the polar hollow cone ($N_{\rm H}^{\rm pol}$), which is calculated along the inner surface of the hollow cone (details in \S\,\ref{sec:cone}). In the following, we describe in detail how the different components used in the models were developed.

\subsubsection{The X-ray source}\label{sec:xsource}

Our model uses as an X-ray source a corona-like component by adopting, for simplicity, the lamp-post scenario (e.g., \citealt{Fabian:2009,emmanoulopoulos+:2014,uttley+:2014}).
To implement such a component we used a sphere that is placed $10\,r_{\rm g}$\footnote{$r_{\rm g}$ is the gravitational radius: $r_{\rm g} = 2GM/c^2$, where $G$ is the gravitational constant, $M_{\rm BH}$ is the black hole mass, and $c$ is the speed of light.} above the SMBH (i.e. the center of the simulation box since there is no physical black hole in the simulation).
A spherical or point source placed at the very center of the simulation would have been fully obscured by the optically thick accretion disk included in our simulation and hence we decided to place it ontop of the SMBH.
Moreover, the system is observed from $0^{\circ}$ to $90^{\circ}$ from the zenith to the plane of the SMBH and therefore we included one X-ray source at $+10\,r_{\rm g}$ without physical motivation for the absence of an X-ray source at $-10\,r_{\rm g}$.
The spherical corona is assumed to have a radius of $6\,r_{\rm g}$ (e.g., \citealt{Chartas:2009,DeMarco:2013}).
To estimate the physical distance we considered the black hole mass of the Circinus galaxy: $\log\left(M_{\rm BH}/M_{\odot}\right) = 6.23$ \citep{Koss+:2017}.
The spatial parameters of the X-ray source are fixed throughout all our simulations.

The next step is to define the spectral properties of the X-ray source.
The photons are generated following a cutoff power law distribution (e.g., \citealt{Zdziarski96,Dadina2008,Ricci+17Catalog}), following the 
\texttt{cutoffpl}\footnote{\url{https://heasarc.gsfc.nasa.gov/xanadu/xspec/manual/node161.html}} model in \textsc{xspec}. 
The simulated photons cover a wide energy range from 0.3 to 300\,keV.
Yet, when collected by \Reflex{} they are divided into two separate files with different binning.
From 0.3 to 10\,keV the original simulation output has \textbf{4\,eV} binning resolution, a decision motivated by the fact that the XRISM/\textit{Resolve} micro-calorimeter (\citealt{xrism:2020}) delivers $\leq 7\,{\rm eV}$ spectral resolution in the energy band $0.3-13\,{\rm keV}$.
Such high spectral resolution comes with computational cost which is translated into noisy spectra. To address this issue, we decided to rebin by a factor of 10 the spectra from 0.3 to 10\,keV, except for the energy regions of the \feka{}, Fe\,K$\beta$ (including the iron edge) and Ni\,K$\alpha$ where the binning remains at 4\,eV.
For the energies above 10\,keV we applied a logarithmic binning with a step size of 0.01.
The cutoff energy of the spectrum is fixed to 200\,keV, corresponding to the median value of nearby AGN \citep{Ricci+:2018}.
The photon index $\Gamma$ varies from 1.6 to 2.4, covering the range typically observed in AGN (e.g., \citealt{Piconcelli2005,Ricci+17Catalog}).
In Table\,\ref{Tab:rxtopo components} we summarize the properties of the X-ray source.

\subsubsection{The dusty torus}\label{sec:torus}

The primary medium responsible for obscuring the central source is a structure that resides after the sublimation zone, usually referred to as "the torus" \citep{Krolik+TORUS:1988ApJ,Jaffe+TorusNGC1068:2004Natur,Mor:2009,Davies+:2015,RamosAlmeida&Ricci:2017Nat}.
Our torus starts at the sublimation radius, which was calculated for a given bolometric luminosity ($L_{\rm{bol}}$), assuming Silicate-type dust particles, using the relation reported by \citet{Mor:2009}:
\begin{equation}
    R_{\rm{sub}}=1.3L^{0.5}_{46} \times \left( \frac{1500\rm{K}}{T_{\rm{sub}}} \right)^{2.6}\,\rm{pc}   
    \label{eq:sublimation}
\end{equation}

where $R_{\rm{sub}}$ and $T_{\rm{sub}}$ are the sublimation radius and temperature, respectively, while 
$L_{46} = L_{\rm{bol}}/10^{46}\rm{\,erg\,s^{-1}}$. 
We adopted the $\rm{2-10\,keV}$ luminosity of the Circinus galaxy ($L_{2-10} = 3\times 10^{42}\,\rm{erg\,s^{-1}}$, \citealp{Arevalo+:2014}), and
converted the X-ray luminosity to bolometric using as bolometric correction $\kappa=20$ \citep{Vasudevan:2009}.

The physical parameters of the torus are set so that the material is neutral and dusty, while the molecular hydrogen fraction is set to $H_2 = 0.3$ (e.g., \citealt{Wilms:2000,Wada+:2009}). For the torus we assumed the \texttt{lpgs} metal abundance \citep{lpgs:2009}, metallicity $Z=1$ and a iron dust depletion of 1.0, which means that all of the Fe atoms are found into dust grains (see \citealt{Ricci&Paltani:2023} for details).
The amount of gas and dust inside the torus is controlled by the volumetric density of the object.
To determine the correct volumetric density of the torus, we divide the equatorial column density, a free parameter, by the diameter of the cross-section of the torus.
This calculation assigns the correct volumetric density in the model via the free parameter $N_{\rm H}^{\rm tor}$.
The parameter range is $\log (N_{\rm H}^{\rm tor}/\rm cm^{-2}) = 22 - 25.5$.
We binned the values of this parameter as follows: from $\log (N_{\rm H}^{\rm tor}/\rm cm^{-2}) = 22$ to 23.3 we considered bins of $\Delta \log (N_{\rm H}^{\rm tor}/\rm cm^{-2})=0.3$, followed by bins of $\Delta \log (N_{\rm H}^{\rm tor}/\rm cm^{-2})=0.2$. These bins have been selected considering the total size of the model grid and the computational cost.

The torus size in the model is regulated by the ratio of the inner radius ($r$) over the distance from the SMBH ($R$).
This ratio is also the $\sin(\phi)=r/R$ which is the angle that corresponds to the covering factor (Fig.\,\ref{fig:rxtopo_cartoon}.
Hence, we introduce the parameter $CF^{\rm tor}= \sin(\phi) = r/R$, which varies from 0.1 to 0.9 with a step size 0.1. The maximum value of 0.9 translates into torus opening angle of $65^{\circ}$.
It should be emphasized that the total sky coverage of the obscuring material may be larger than the covering factor of the torus, as the polar component also contributes to it. (see, \S\,\ref{sec:cone}). 
The properties of the torus adopted in our simulations can be found in Table\,\ref{Tab:rxtopo components}, while the column density profile of this component is shown in Fig.\,\ref{fig:rxtopo_dens_profile} for two different covering factors.

\subsubsection{The polar cone}\label{sec:cone}

\begin{figure*}[ht!]
    \centering
    \includegraphics[width=0.45\textwidth]{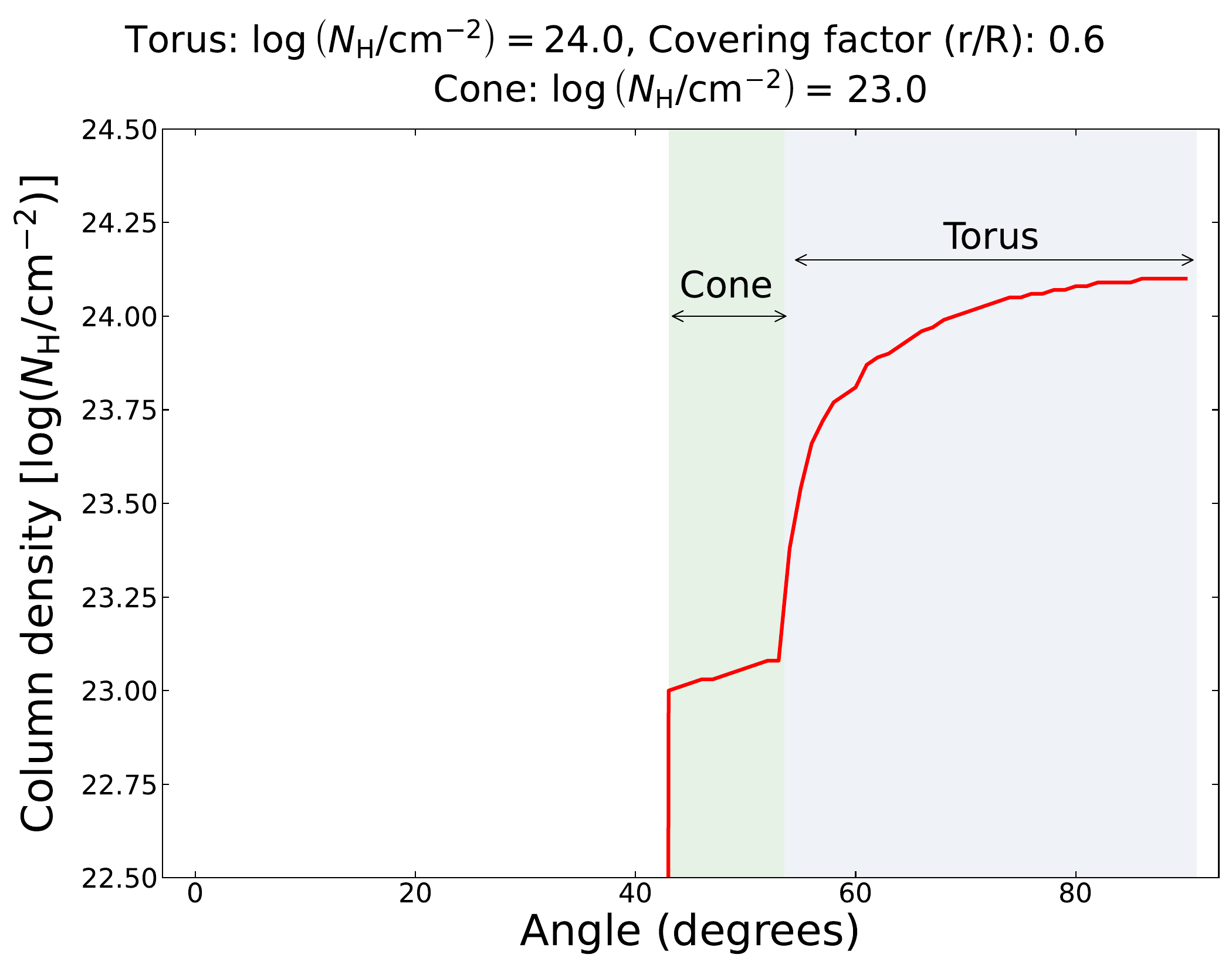}
    \hfill
    \includegraphics[width=0.45\textwidth]{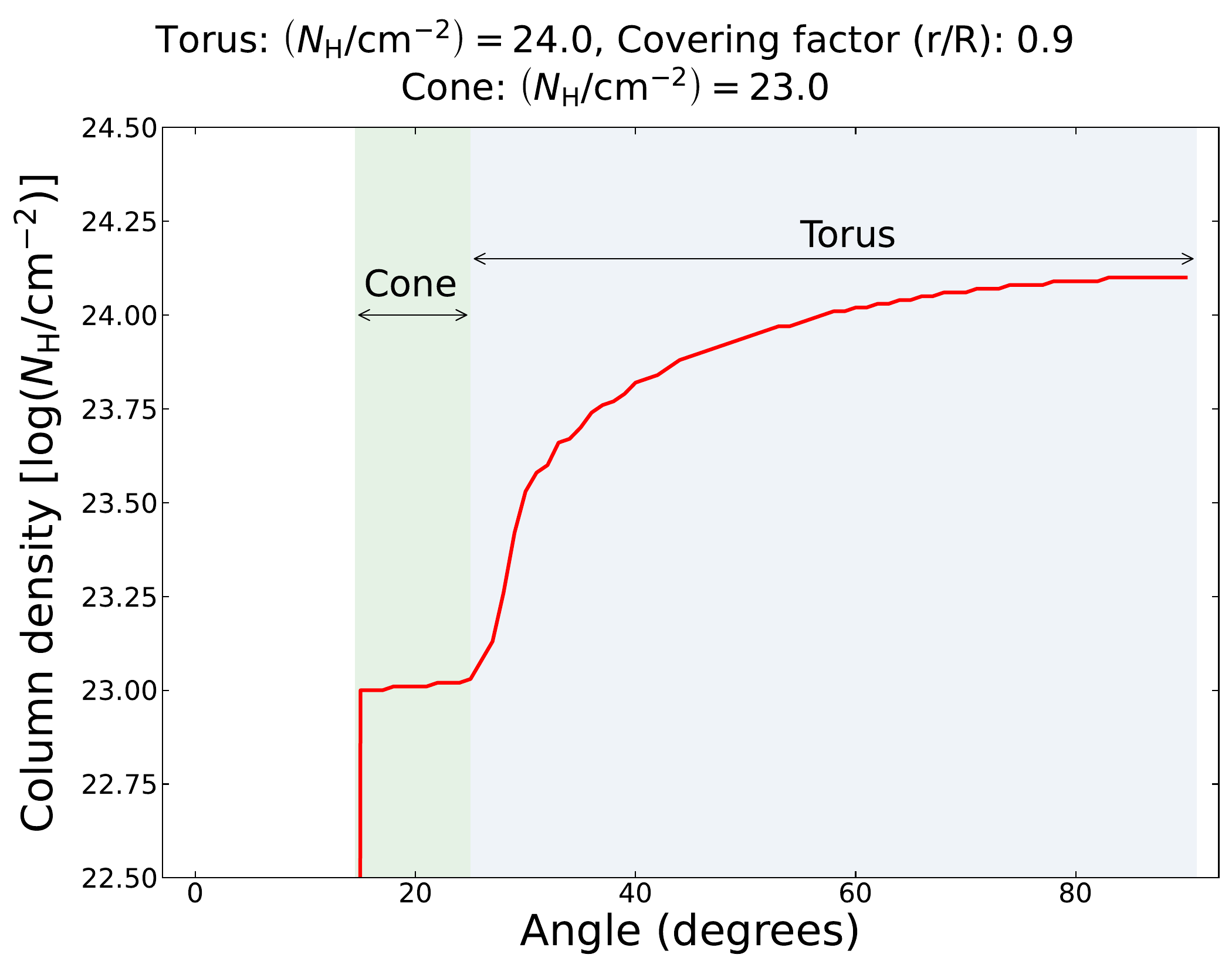}
    \caption{Two examples of column density profile for the \rxtopo{} model, considering two different torus configurations. In both panels we show the line-of-sight column density varies for different angles. The shaded areas that are also noted by the arrows mark the angle range in which the cone or torus intercept the photons that originate from the center. In the left panel we considered a torus with $CF^{\rm tor}= (r/R) = 0.6$, whereas the right panel shows a torus with $CF^{\rm tor}= 0.9$.}
    \label{fig:rxtopo_dens_profile}
\end{figure*}

Infrared observations of local AGN suggest the presence of an extended region that can reach hundreds of parsecs, composed of warm dust (e.g. \citealt{Stalevski+:2017,Asmus:2019,GATOSII:2021A&A}).
In the \rxtopo{} model, we implement this dusty structure by adopting a hollow cone geometry, similar to that which has been shown to effectively reproduce the IR emission of the Circinus galaxy \citep{Stalevski+:2017,Stalevski+:2019}. 
In Fig.\,\ref{fig:rxtopo_cartoon} we illustrate the cone component, while the column density profile of the \rxtopo{} model is shown in Fig.\,\ref{fig:rxtopo_dens_profile}.
The hollow cone has the angular width set to $10^{\circ}$ the findings of \citet{Stalevski+:2017} for the Circinus galaxy.
The opening angle of the cone, or in other words, the angle at which the cone starts intercepting the X-ray source, is controlled by the torus. In our model the torus is assumed to collimate the polar cone, since this structure is thought to be associated with material that has been expelled from the dusty torus (e.g., \citealt{Venanzi:2020}).
We use the opening angle of the torus ($\phi$, see for details \S\,\ref{sec:torus}), and we set the cone to have a maximum angle $\xi_{\rm max}^{\rm pol} =\phi + 1^{\circ}$, to ensure that the two components do not overlap.
The opening angle of the cone is therefore $\xi^{\rm pol}_{\rm o}= \xi_{\rm max}^{\rm pol} + 10^{\circ}$.

The material in this component is set to be neutral with metallicity $Z = 1$ and molecular hydrogen fraction $H_2 = 0.3$, similar to the torus, since they reside both at similar distances from the SMBH.
Similarly to what was done for the torus, the metal abundance adopted is \texttt{lpgs} \citep{lpgs:2009}.
The polar cone is placed above the sublimation radius since it is assumed to be dusty. Therefore, we set the base of the geometrical object at $R_{\rm sub}$ (see equation\,\ref{eq:sublimation}), as it was done for the torus.
The material the eventually settles in the polar cone originates from inner parts of the system and therefore it is natural to assume that there should be another "component" that lies between the innermost regions and the polar cone.
Yet, the main motivation for the polar cone is to account for the extended emission observed in the IR, and therefore we decided to include it in the model. Moreover, the X-ray spectrum is affected by the column density across the line of sight and hence the free parameter $N_{\rm H}^{\rm pol}$ accomodates any additional neutral material that could be present in the system.
Note, that ionized winds are present in many cases but they are not part of this study and therefore ionized material and high velocity material will be included in future versions of the models when they are available in \Reflex{} (e.g., \citealt{Psaradaki_outflows:2024arXiv}).
The dust depletion is also set to 1.0 (see \S\,\ref{sec:torus} and \citealp{Ricci&Paltani:2023}), since it is expected that the dust grains will form rapidly after passing the sublimation zone due to the outflowing material (e.g., \citealt{Mehdipour_dust:2018AnA}).
The outer boundary of the conical structure is set to 40\,pc.
Although there is no clear end point to the upper boundary of the cone and the material is expected to be distributed even further away (e.g., \citealt{Asmus:2019} and more recently \citealt{Haidar+JWSTcone:2024ARXIV}), our motivation is to include the most dense circumnuclear region, which can influence the X-ray photon field that emerges from the corona.
The column density of the hollow cone is a free parameter and is calculated along the innermost surface of the cone.
It spans a range of $\log (N_{\rm H}^{\rm pol}/\rm cm^{-2}) = 22.0 - 24.0$ (e.g., \citealt{Asmus:2019,Venanzi:2020}), with a step size of $\Delta \log (N_{\rm H}^{\rm pol}/{\rm cm^{-2}}) =0.25$.
In Appendix\,\ref{app:cone_vs_torus} we demonstrate the influence of the hollow polar cone component on the reprocessed spectrum.
We report the properties of the cone in Table\,\ref{Tab:rxtopo components} and in Fig.\,\ref{fig:rxtopo_dens_profile} the effect of the hollow polar cone on the line-of-sight column density is highlighted.

\begin{table}
\centering
\caption{Properties of the components used in the \rxtopo{} model. The free parameters of the model are shown in boldface. Note that column densities are expressed in units of $[\log \left(N_{\rm H}/{\rm cm^{-2}}\right)]$. More details are included in \S\,\ref{sec:rxtopo}}
\begin{tabular}{c c}
\hline
\hline
\multicolumn{2}{c}{\multirow{2}{*}{\textbf{\large{X-ray source}}}} \\
\\
\Reflex{} element & sphere \\
\textbf{$\mathbf{\Gamma}$} & \textbf{1.6 - 2.4} \\
Ec [keV] & 200 \\
Band & 0.3 - 300\,keV \\
Binning (0.3 - 10\,keV) & linear $\Delta E=20$ eV  \\
Binning (10 - 300\,keV) & $\Delta \log(E/\rm eV)=0.01$ \\
\textbf{Observing angle} & $\mathbf{0^{\circ}-90^{\circ}}$ [step $2^{\circ}$]\\
\hline
\multicolumn{2}{c}{\multirow{2}{*}{\textbf{\large{Torus}}}} \\
\\
\Reflex{} element & torus \\
State & neutral \\
$\rm H_2$, $Z$ & 0.3, 1 \\
Dust & 1.0 \\
\textbf{Column Density} & \textbf{22.0 - 25.5} \\
{$\mathbf{CF^{\rm tor}\,(r/R)}$} & \textbf{0.1 - 0.9} \\
\hline
\multicolumn{2}{c}{\multirow{2}{*}{\textbf{\large{Hollow Cone}}}} \\
\\
\Reflex{} element & Cone \\
State & neutral \\
$\rm H_2$, $Z$ & 0.4, 1 \\
Dust & 1.0 \\
Angle (width) & $\rm 10^{\circ}$ \\
\textbf{Column Density} & \textbf{22.0 - 24.0} \\
\hline
\end{tabular}
\label{Tab:rxtopo components}
\end{table}

\subsection{The RXagn1 model}\label{sec:rxagn}

The second new model presented here is \rxagn{}, an extension of the \rxtopo{} model that incorporates additional components to account for all relevant AGN structures. In addition to the torus (\S\ref{sec:torus}) and the polar medium (\S\ref{sec:cone}), \rxagn{} includes the accretion disk (\S,\ref{sec:disk}) and the broad line region (\S,\ref{sec:blr}), both of which are well-established components of the circumnuclear environments around SMBHs (e.g., \citealt{Netzer_UniModelreview:2015}).

The \rxagn{} model has the same configuration for parameters shared with \rxtopo{}. In other words, the torus has as free parameters the column density ($N_{\rm H}^{\rm tor}$) and the covering factor ($CF^{\rm tor}$). However, we decided to set a minimum covering factor of the torus $CF^{\rm tor}= 0.2$. The motivation behind this change in the range of the $CF^{\rm tor}$ is the presence of the broad line region (BLR). Even though the two elements (BLR, Torus) are not linked within our models we consider unphysical the combination of a very small torus along with a bigger BLR. Hence, the range of the covering factor of the torus is $CF^{\rm tor}=0.2-0.9$. 
In the case of the polar hollow cone the column density across the inner surface is a free parameter as presented in \S\ref{sec:cone}.
Finally, similar to the \rxtopo{} model the photon index and the observing angle are free to vary (see \S\ref{sec:rxtopo} for details).
The two new elements included, BLR and accretion disk, have their parameters fixed and are summarized in Table\,\ref{Tab:rxagn_components}.
The \rxagn{} model is illustrated in Fig.\,\ref{fig:rxagn_cartoon}. To demonstrate the influence of the BLR, Fig.\,\ref{fig:rxagn_dens_profile} shows examples of the line-of-sight column density for two different tori configurations.

\subsubsection{The accretion disk}\label{sec:disk}

The accretion disk (AD) is the structure that is expected to "feed" the SMBH, and it is widely thought to be responsible for the strong nuclear emission observed in the optical and UV. To model the AD we adopted the Shakura-Sunyaev (SS73) model \citep{Shakura&Sunyaev:1973}, which considers a geometrically-thin, optically-thick accretion disk.
We set the disk to start at $10\,r_{\rm g}$ from the center and extend up to $1000\,r_{\rm g}$, to be consistent with recent studies of optical variability (e.g. \citealp{McHardy+:2023}).
We set the metallicity to unity, and used the \texttt{lpgs} abundance for consistency with the other objects.
Due to the high temperatures in the disk, the hydrogen and helium are fully ionized and the medium dust-free.
Finally the volumetric density is fixed to $n_{\rm H} = 10^{12}\,{\rm cm^{-3}}$ (e.g., \citealt{Garcia+:2013}).
The current version of our model does not incorporate any general relativity effect. In Table\,\ref{Tab:rxagn_components} we summarize the properties of the disk.

\subsubsection{The broad line region}\label{sec:blr}

\begin{figure*}[ht!]
    \centering
    \begin{subfigure}
        \centering
        \includegraphics[width=0.45\textwidth]{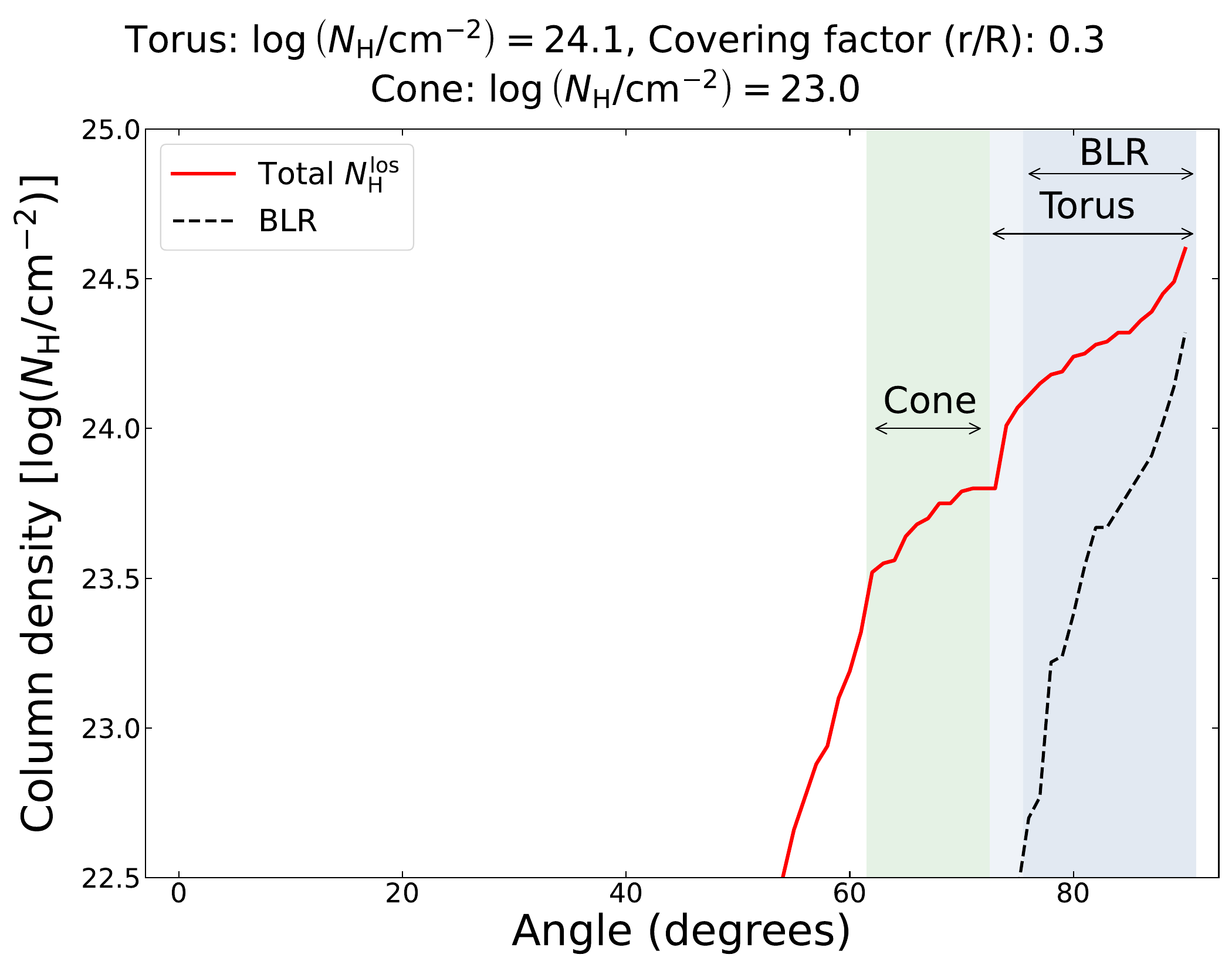}
    \end{subfigure}
    \begin{subfigure}
        \centering
        \includegraphics[width=0.45\textwidth]{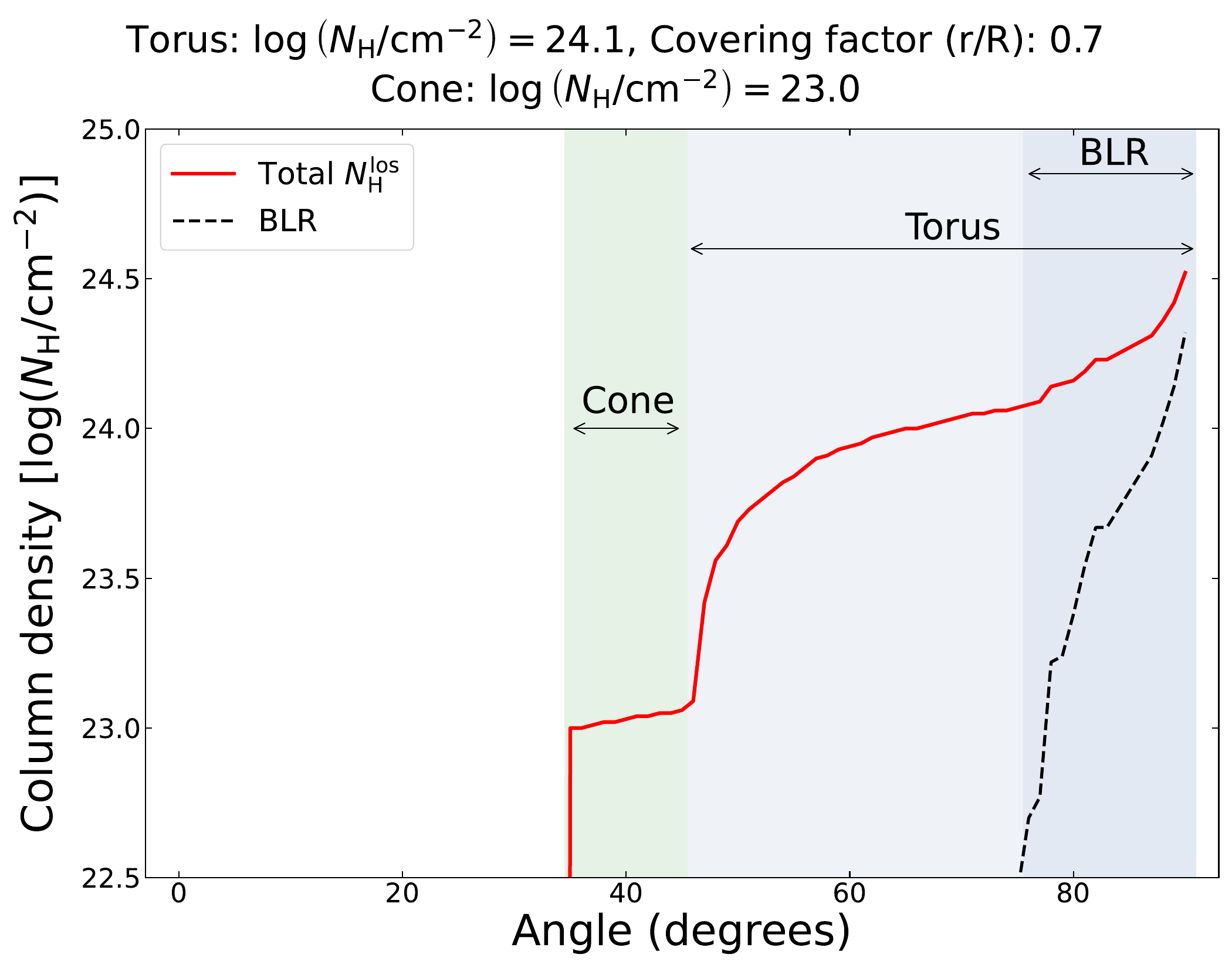}
    \end{subfigure}
    \caption{The line-of-sight density profile of the \rxagn{} model. The solid line represents the total line-of-sight column density whereas the dashed line illustrates the contribution of the broad line region (BLR). Left: The density profile of the \rxagn{} model for $CF^{\rm tor}=0.3$. Right: The density profile of a larger torus with $CF^{\rm tor}=0.7$.}
\label{fig:rxagn_dens_profile}
\end{figure*}

The second additional object included in the \rxagn{} model is the broad line region (BLR). The BLR is the region where broad emission lines are produced, which implies the presence of gas that revolves at velocities that can reach up to a few hundreds thousands of km/s.
The current version of \rxagn{} model does not include any velocity broadening associated with the motion of the BLR clouds. It will be implemented in a future version of \Reflex{}.
Yet, the BLR can provide additional obscuration for high inclination angles (see Fig.\,\ref{fig:rxagn_dens_profile}).
We assume that the BLR extends from the end of the AD up to the sublimation zone where the torus starts.
To model the BLR we used a flare disk model (e.g., \citeauthor{GravityCollab:2018Natur} \citeyear{GravityCollab:2018Natur}, \citeyear{GravityColab:2020}).
To develop this model, we used the annulus geometry provided by \Reflex{}. A series of 30 consecutive annuli were generated. These annuli vary in width (i.e. inner and outer radius from the center of the simulation) and height.
Using this method, we develop a model for the BLR, where its center is associated with the luminosity-weighted distance as described by \citet{Kaspi:2005}.
In our model the height of the BLR annuli increases logarithmically.
The volumetric density spans from $n_{\rm H}=10^9\,{\rm cm^{-3}}$ to $10^6\,{\rm cm^{-3}}$, which is consistent with a medium that connects the denser disk with the lower density torus (e.g., \citealp{Ferland+:1992,Goad+:2012}).
In Fig.\,\ref{fig:rxagn_dens_profile} the column density profile of \rxagn{} model is illustrated including the BLR.
In our simulations, the BLR is dust-free since it ends at the sublimation zone.
The gas is assumed to be neutral and the molecular hydrogen fraction is set to 20\%.
We use the \texttt{lpgs} metal abundance, and the metallicity is fixed to unity.
In Table\,\ref{Tab:rxagn_components} we summarize the properties of the BLR.


\begin{figure}[ht!]
    \centering
    \includegraphics[width=0.9\columnwidth]{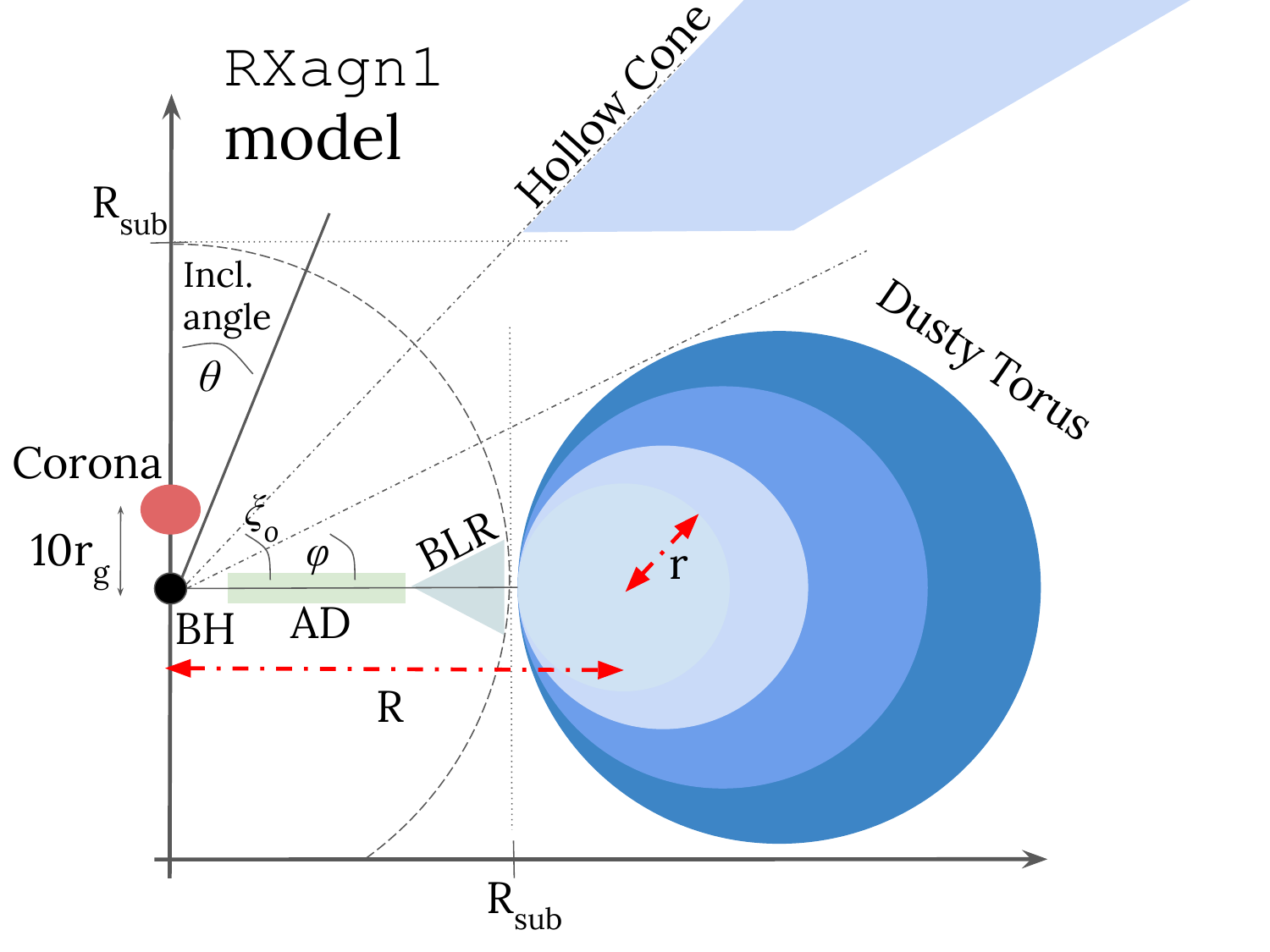}
    \caption{Cartoon illustrating the configuration of the \rxagn{} model. This model is an extension of the \rxtopo{} model, with the addition of an accretion disk (AD) and a broad line region (BLR). The dash-dotted line (red) depicts the R and r that are used to calculate the covering factor of the torus. The objects are not scaled. The angle $\theta$ is the observing angle of the system. $\phi$ is the opening angle of the torus and $\xi_{\circ}$ is the maximum angle of the polar hollow cone.}
    \label{fig:rxagn_cartoon}
\end{figure}

\begin{table}[]
\centering
\caption{Properties of the extra components used in the \rxagn{} model. All the parameters of the accretion disk and BLR are fixed. The \rxagn{} model includes the components of Table\,\ref{Tab:rxtopo components} too.}
\begin{tabular}{c c}
\hline
\hline
\multicolumn{2}{c}{\multirow{2}{*}{\textbf{\large{Accretion disk}}}} \\
\\
\Reflex{} element & disk \\
State & H,He ionized \\
$H_2$ & 0 \\
$Z$ & 1 \\
Volumetric density ($n_{\rm H}$) & $10^{12}\,{\rm cm^{-3}}$ \\
Dust & 0 \\
\hline
\multicolumn{2}{c}{\multirow{2}{*}{\textbf{\large{Broad line region}}}} \\
\\
\Reflex{} element & complex of annuli \\
State & neutral \\
$H_2$ & 0.2 \\
$Z$ & 1 \\
Volumetric density ($n_{\rm H}$) & $10^{9}-10^{6}\,{\rm cm^{-3}}$ \\
Equatorial column density & $N_{\rm H}\sim 2.5\times10^{24}\,{\rm cm^{-2}}$\\
Dust & 0 \\
\hline
\end{tabular}
\label{Tab:rxagn_components}
\end{table}

\section{The models}\label{sec:comparison}

\subsection{The spectral models}\label{sec:different_components}

As described in \S\,\ref{sec:reflex} \Reflex{} gives the user the freedom to select the photons collected according to the type of interactions they have undergone.
For example, continuum photons can be distinguished from scattered or fluorescent photons.
We have used this option when transforming the simulation output into spectral model files, providing five different model components of both the \rxtopo{} and \rxagn{} model (see details about the simulations in \S\,\ref{sec:sim_setup}).

\begin{figure}[ht!]
    \centering
    \includegraphics[width=0.9\columnwidth]{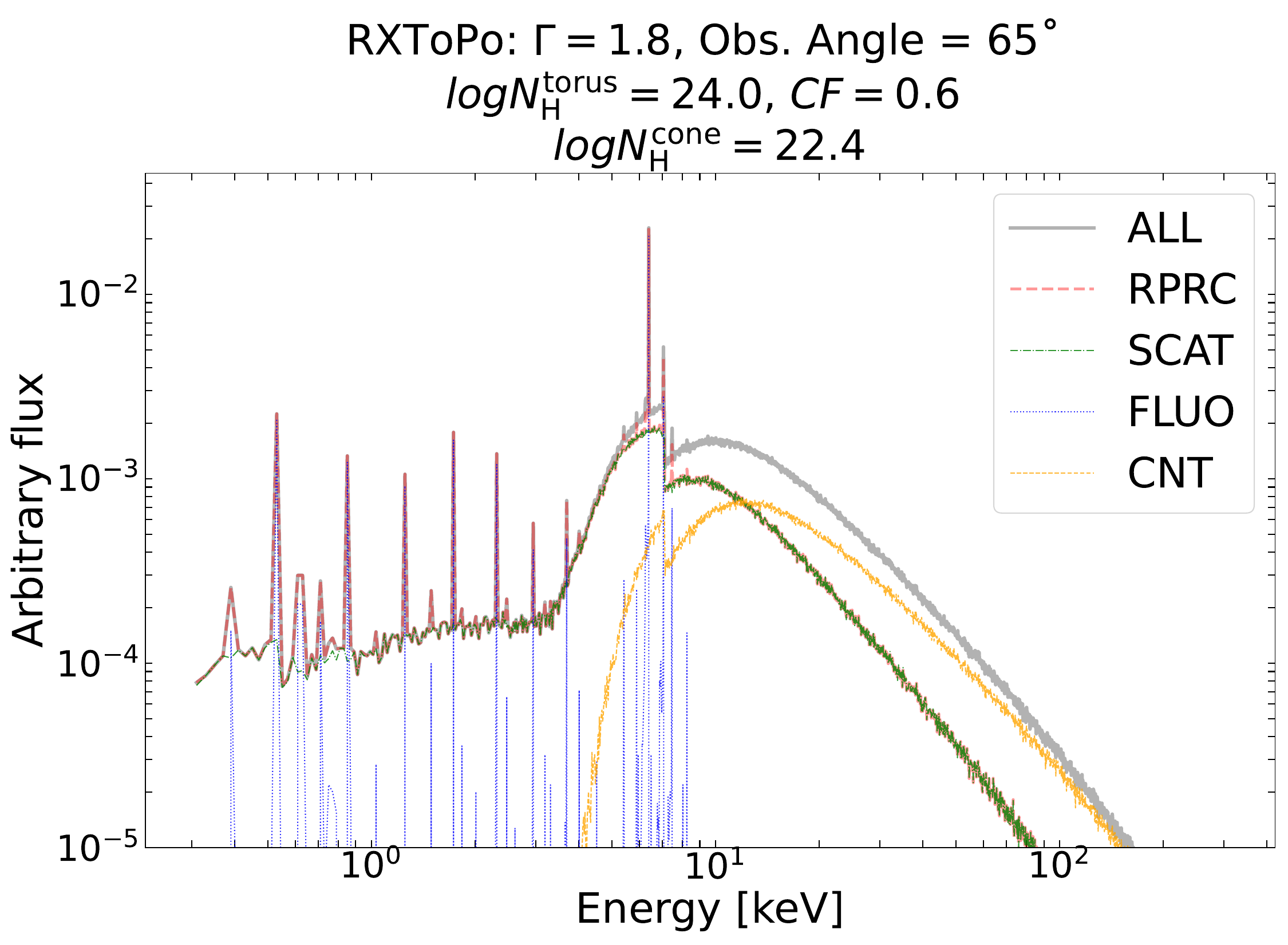}
    \caption{Different options available for the \rxtopo{} and \rxagn{} models as described in \S\,\ref{sec:different_components}.
    The different lines represent : all the photons collected (ALL, grey solid line), reprocessed photons (either by scattering or fluorescence; RPRC, dashed red line), scattered photons (SCAT, dashed green line), fluorescent photons (FLUO, solid blue line), continuum photons (CNT, orange dashed line) , Here an example of the \rxtopo{} model with the following parameters: Photon index $\Gamma = 1.8$, Obs. Angle = $65^{\circ}$, Torus $\log (N_{\rm H}^{\rm tor}/{\rm cm^{-2}})=24.0$, $CF^{\rm tor}=0.6$ and Cone $\log (N_{\rm H}^{\rm pol}/{\rm cm^{-2}})=22.4$.}
\label{fig:different_flavors}
\end{figure}

The first option given to the user is the compilation of all the photons collected at the end of the simulation.
We mark this as "ALL" to better illustrate that it consists of both continuum and reprocessed photons.
The second option includes only photons that have been reprocessed by the medium.
This reprocessed component is noted as "RPRC" and in other words is the spectrum subtracted continuum.
Next, there is an option that consists of reprocessed photons excluding any kind of fluorescence.
This option is named "SCAT".
The fluorescent lines are also available as separate model file, which is marked as "FLUO".
Finally, we provide a spectral component that is the sum of continuum photons that originate from the central source and did not interact with the medium, whatsoever.
It is called "CNT" for continuum.
Overall, the structure: ALL = CNT + RPRC, where RPRC = SCAT + FLUO.
\textbf{In Fig.\,\ref{fig:different_flavors} is presented an example featuring all the different components described above.}

\begin{figure*}[ht!]
    \centering
    \includegraphics[width=0.9\textwidth]{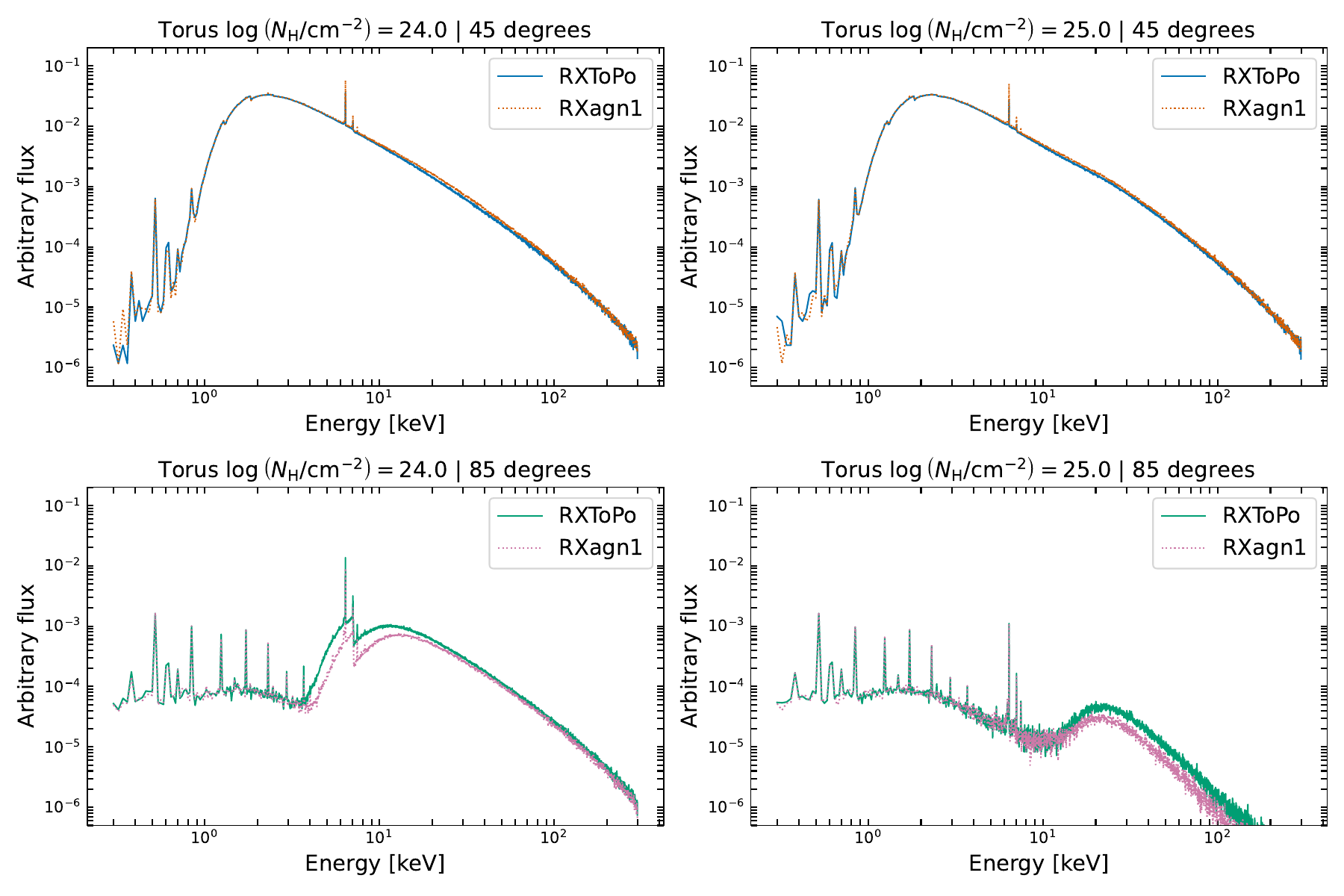}
    \caption{The four panels illustrate a few examples of the simulated spectra. In every panel are presented both the \rxtopo{} and \rxagn{} model. The top row shows two different versions of the torus with $\log (N_{\rm H}^{\rm tor}/{\rm cm^{-2}})=24.0$(left),\,$25.0$(right), $CF^{\rm tor}=0.6$ and Cone $\log (N_{\rm H}^{\rm pol}/{\rm cm^{-2}})=22.4$ observed at 45 degrees. The bottom row is the same configuration but observed at 85 degrees.}
\label{fig:comparison_rxmodels}
\end{figure*}

These five component model types provide users with the flexibility to select the one that is appropriate according to the source being studied.
In Fig.\,\ref{fig:comparison_rxmodels} we present a few examples of the \rxtopo{} and \rxagn{}.
In the figures, we have selected two different observing angles, $45^{\circ}$ and $85^{\circ}$, as well as two torus column densities $\log \left(N_{\rm H}^{\rm tor}/{\rm cm^{-2}}\right)=24.0,\,25.0$.
The photon index is $\Gamma=1.8$, the torus covering factor is $CF^{\rm tor}=0.6$, while the polar cone has column density $\log \left(N_{\rm H}^{\rm pol}/{\rm cm^{-2}}\right)=22.4$.
In the top panels where the observing angle does not intercept the BLR the two modeled spectra are similar, yet in the bottom panels where the BLR intercepts the line of sight the \rxagn{} model shows higher absorption.

\subsection{Application on NGC 424}\label{sec:fitNGC424}

Having developed these two new models, \rxtopo{} and \rxagn{}, we are interested in applying both models to test their capabilities.
We applied both models to NGC\,424, one of the closest Compton-thick (CT) AGN detected by Swift/BAT \citep{Ricci+:2015}, at a luminosity distance of $d_{\rm L} = 45.6\,{\rm Mpc}$ ($z = 0.011761$). NGC\,424 was confirmed to be CT by \citealt{Matt+:2000BeppoCT} (Tololo\,0109$-$383) using \textit{BeppoSAX} observations. They found that the central source is obscured by cold material that exhibits a column density of $2\times10^{24}\,{\rm cm^{-2}}$.
Later, \citealt{Marinucci+:2011ngc424} studied thoroughly the same source using \textit{XMM-Newton} Reflection Grating Spectrometer (RGS) and EPIC-pn and found that the soft part ($< 2$\,keV) of the spectrum shows several emission lines from ionized material, and the harder ($> 2$\,keV) part shows strong reprocessed features like iron lines K$\alpha$ and K$\beta$ as well as Ni\,K$\alpha$.
\citealt{Balokovic+:2014_CTsources} studied NGC\,424 along with other CT sources by using \textit{NuSTAR}, \textit{Swift}/XRT and \textit{XMM-Newton} data. Their analysis further supports the presence of a dominant reflection component in the X-ray spectrum, which makes it a good candidate to test our newly developed models.
The obscured nature of NGC\,424 was also confirmed by \citet{Marchesi_CTnustar:2019ApJ} where they applied two different models -- \texttt{MyTorus}, \texttt{borus02} -- on a compilation of X-ray data and in both cases the line-of-sight column density is $> 10^{23}\,{\rm cm^{-2}}$.
They have also tried the \texttt{borus02} model having the covering factor ($fc$) free that gave photon index $\sim 1.8$ for $fc=0.4$.
In a more recent work, \citet{Tanimoto_2022} applied \textbf{the clumpy torus model \texttt{XClumpy} \citep{Tanimoto:2019} }in a big sample of obscured sources.
For NGC\,424 they found that the line-of-sight is $\sim 0.84\times 10^{24}\,{\rm cm^{-2}}$ for equatorial column density $\sim 1.5\times 10^{24}\,{\rm cm^{-2}}$.
The photon index retrieved was $\Gamma \sim 1.7$.

Moreover, \citealt{Paltani&Ricci:2017} already tested the \rxtorus{} model on the same source as a benchmark for the model.

We use here both \textit{XMM-Newton}\,EPIC data and \textit{NuSTAR}\,FPMA/FPMB to benchmark the \rxtopo{} and \rxagn{} models. We analyzed the \textit{XMM-Newton}\,EPIC spectrum of the source using an 8.2\,ks observation (ID\,0002942301, PI M. Guainazzi) along with a 15.5\,ks \textit{NuSTAR} observation (ID\,60061007002), which was performed as part of the campaign to follow up AGN detected by Swift/BAT.
The observations are the same as those used in \citet{Paltani&Ricci:2017}.
We followed typical procedures for data reduction and spectral extraction, and details can be found in \citealt{Paltani&Ricci:2017}.

\subsubsection{Spectral fitting}\label{sec:ngc424specfit}

For the X-ray spectral fitting, throughout this study we use \textsc{xspec} \citep{XSPEC}.
We set the cosmic abundance to \texttt{lpgs} to be consistent with our models.
Moreover, all spectra have been rebinned to contain at least 25 counts in each energy bin to apply $\chi^2$ statistics.
All errors reported are calculated in the confidence region 90\%.
For both models, we set the Galactic absorption \textbf{to $N_{\rm H\,gal} = 1.59\times 10^{20}\,{\rm cm^{-2}}$,} as derived by \citealt{HI4PI:NH:2016A&A}.

The first reflection model that we applied is \rxtopo{}.
The \rxtopo{} model consists of a torus that is in the plane of the accretion disk, with a hollow polar cone perpendicular to it, as described in \S\,\ref{sec:rxtopo}.
The free parameters of \rxtopo{} are: the observing angle of the system ($\theta_{\circ}$), the photon index $\Gamma$ of the incident power-law continuum, the equatorial hydrogen column density of the torus ($N_{\rm H}^{\rm tor}$), the covering factor of the torus defined as the ratio between the major axis (R) and minor axis (r) of the torus ($CF^{\rm tor}=r/R$) and the hydrogen column density of the hollow cone measured along the inner surface of the cone ($N_{\rm H}^{\rm pol}$).
The softer part of the spectrum shows several spectral features that can be attributed to ionized gas, and therefore we
included an \texttt{apec}\footnote{\url{https://heasarc.gsfc.nasa.gov/xanadu/xspec/manual/node134.html}} model.
After several tests we decided to add a second \texttt{apec} component to further improve the fit of the soft X-rays.
The free parameter of the two \texttt{apec} is the temperature of the plasma in keV, as well as the normalization. The metal abundance is fixed to the default value (1.0). 
Finally, a Gaussian emission line is included to reproduce ionized iron ($\rm Fe\,XXVI$), with width fixed at 10\,eV, using the \texttt{zgauss} model.
The final model in the \textsc{xspec} notation is:
\begin{equation}
    tbabs\times(rxtopo_{\rm all} + apec_1 + apec_2 + zgauss)
\end{equation}

\begin{figure}[ht!]
    \centering
    \includegraphics[width=0.9\columnwidth]{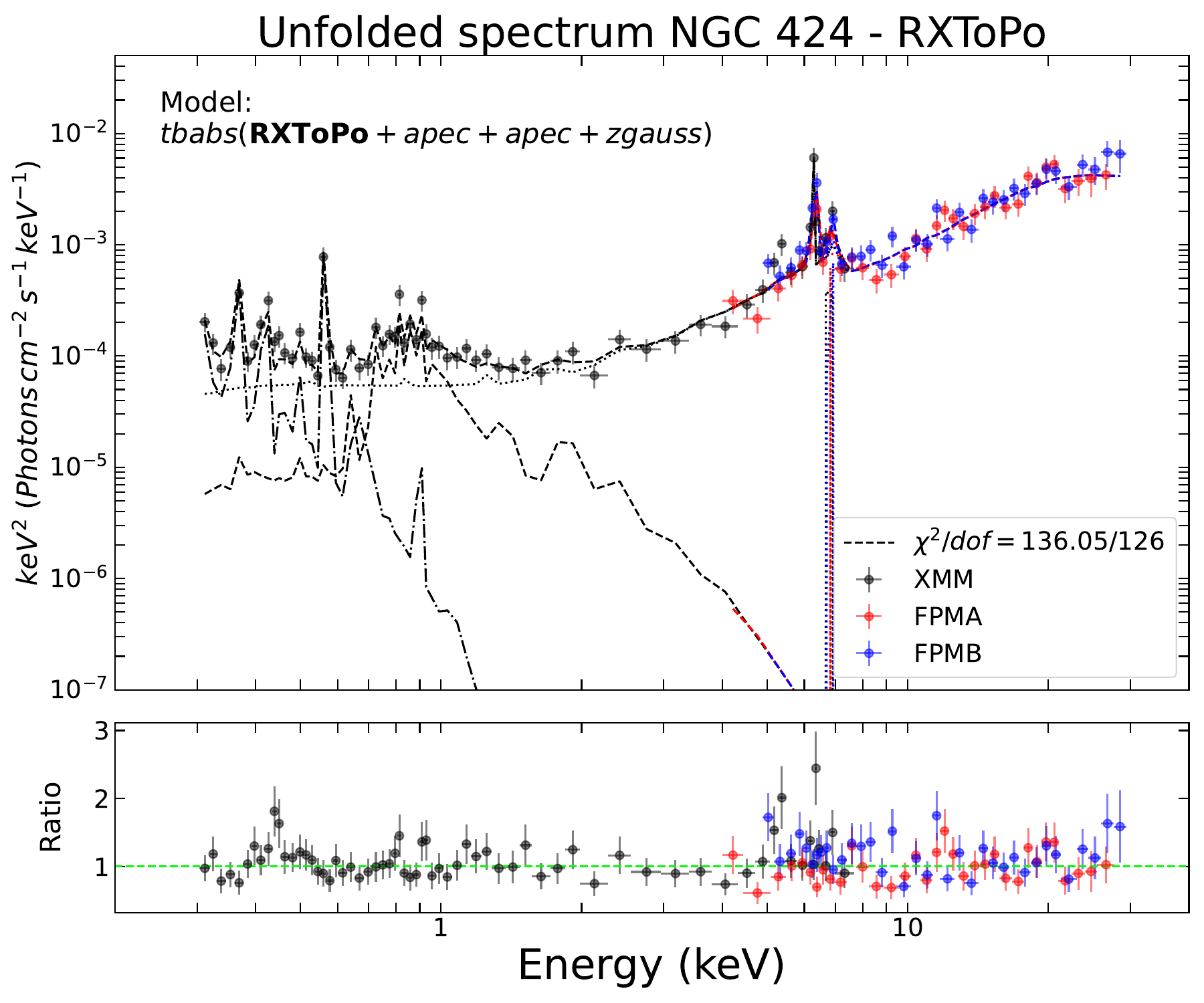}
    \caption{The \textit{XMM-Newton}\,EPIC (black points) and \textit{NuSTAR} (red and blue points) data for NGC\,424 along with the best fit model. The individual components are also present. The bottom panel shows the ratio between the data and the model.}
\label{fig:rxtopo_fit}
\end{figure}

The model fits well the data, achieving $\chi^2 /dof = 136.05/126$, with null-hypothesis probability $p_{\rm value} = 0.255$.
The unfolded spectrum and the ratio between the best fit and the data are presented in Fig.\,\ref{fig:rxtopo_fit}).
We find that the photon index is $\Gamma = 1.80^{+0.01}_{-0.01}$, and the observing angle $79.0^{\circ \, +1}_{\;\; -1}$.
The equatorial hydrogen column density of the torus is $\log \left(N_{\rm H}^{\rm tor}/{\rm cm^{-2}}\right)=24.48^{+0.03}_{-0.08}$, with a covering factor $CF^{\rm tor} = 0.66^{+0.02}_{-0.01}$, while the polar cone has $\log \left(N_{\rm H}^{\rm pol}/{\rm cm^{-2}}\right)=23.19^{+0.04}_{-0.06}$.
In order to obtain the line-of-sight column density the simpler approach is to calculate it using the formula:
\begin{equation}
    N_{\rm H}^{\rm los} = N_{\rm H}^{\rm tor}\left(1 - \left(\frac{1}{CF^{\rm tor}}\right)^2  cos^2(\theta_{\circ})\right)^{1/2}
\end{equation}

which, for the parameters retrieved from the fit, results in $\log \left(N_{\rm H}^{\rm los}/{\rm cm^{-2}}\right)=24.46$ or $2.89 \times 10^{24}\,{\rm cm^{-2}}$. This approach is valid only when the line-of-sight interecepts the dusty torus.
In our case, $CF^{\rm tor} = 0.66$ means that the torus extends up to $\sim 42^{\circ}$ from the horizon ($\arcsin(CF^{\rm tor})$), while the observing angle is $79^{\circ}$, which corresponds to just $11^{\circ}$ from the horizon.
\Reflex{} also provides an alternative method for calculating $N_{\rm H}^{\rm los}$ in any geometry and configuration.
We can in fact use the \texttt{REFLEXINO} tool, which is similar to that used to generate Fig.\,\ref{fig:rxtopo_dens_profile}, \ref{fig:rxagn_dens_profile}. Details on this approach, and how to build this simulation type are found in Appendix\,\ref{app:reflexino}.
Consistent with what was expected, the value found with this method is $\log \left(N_{\rm H}^{\rm los}/{\rm cm^{-2}}\right)=24.46$.
The spectral fitting results are summarized in Table\,\ref{tab:spectral_fitting}.

Next, we apply the \rxagn{} model, which is an extension of \rxtopo{} that includes the BLR and the accretion disk (\S\,\ref{sec:different_components}).
The \rxagn{} model has the same free parameters ($\theta_{\circ}$, $\Gamma$, $N_{\rm H}^{\rm tor}$, $CF^{\rm tor}=r/R$, $N_{\rm H}^{\rm pol}$) as the \rxtopo{}.
We include two \texttt{apec} components, leaving the plasma temperature free to vary with their abundance fixed to the default value; unity.
Finally, as done for the \rxtopo{} model, a Gaussian line is included, with the width set to $\sigma = 10$\,eV.
The model is:
\begin{equation}
    tbabs\times(rxagn1_{\rm all} + apec_1 + apec_2 + zgauss)
\end{equation}

\begin{figure}[ht!]
    \centering
    \includegraphics[width=0.9\columnwidth]{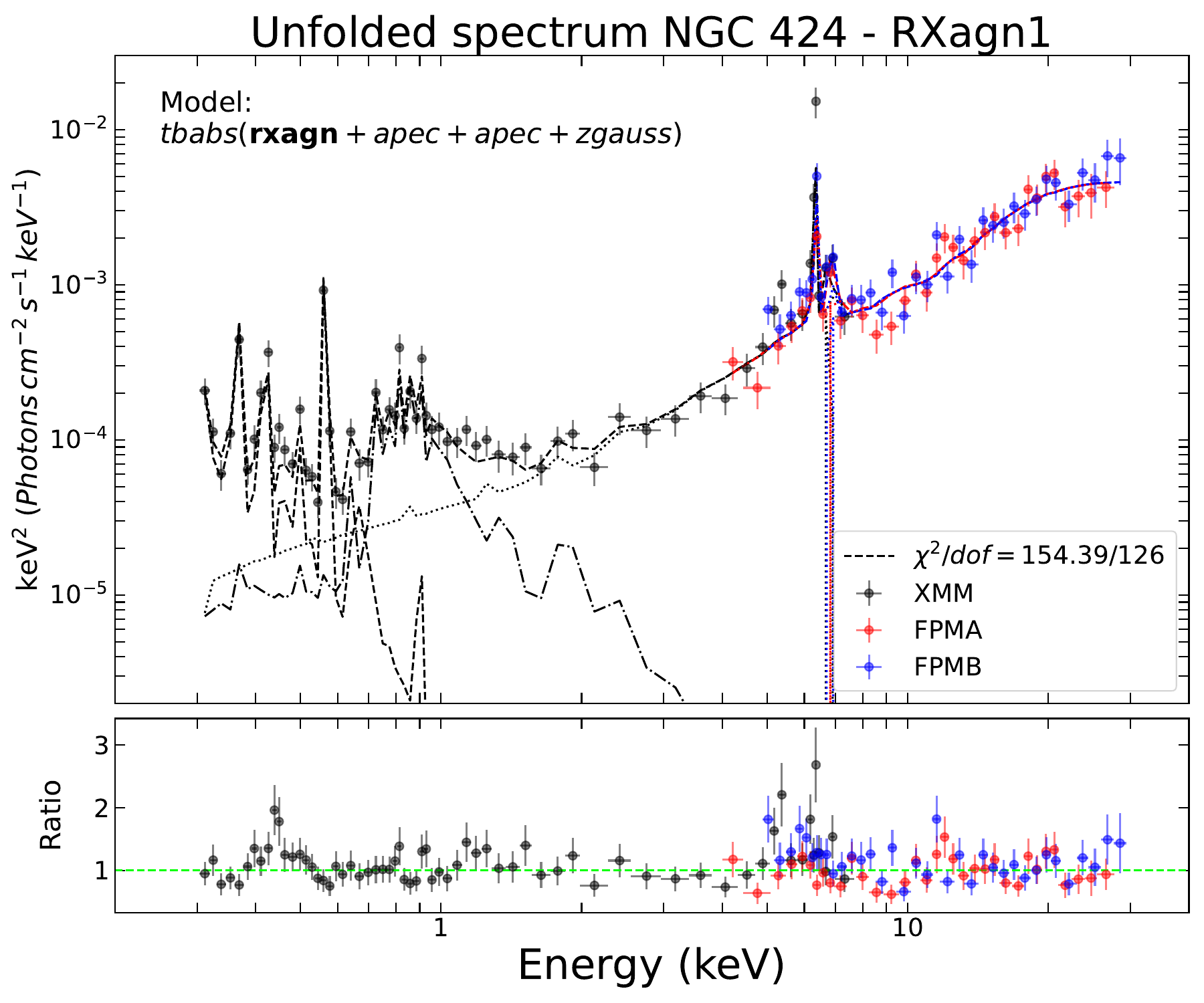}
    \caption{The \textit{XMM-Newton}\,EPIC (black points) and \textit{NuSTAR} (red and blue points) data for NGC\,424 along with the best fit model. The individual components are also present. The bottom panel shows the ratio between the data and the model.}
\label{fig:rxagn_fit}
\end{figure}

This model also provides a good fit ($\chi^2/dof= 154.39/126$) with null-hypothesis probability $p_{\rm value} = 0.044$, and is presented in Fig.\,\ref{fig:rxagn_fit}. We find that the photon index is $1.78^{+0.01}_{-0.01}$ and the observing angle $74.9^{\circ\,+0.4}_{\;\;-2.0}$.
The column density of the torus is $23.90^{+0.01}_{-0.01}$ for covering factor $\geq 0.90^{+u}_{-0.01}$. The polar cone shows a column density higher than that of the \rxtopo{} model [$\log \left(N_{\rm H}^{\rm tor}/{\rm cm^{-2}}\right)=23.80^{+0.01}_{-0.01}$].
In order to find the line-of-sight column density for the current configuration, we use the \texttt{REFLEXINO} tool (see Appendix\,\ref{app:reflexino}).
We find that the line-of-sight column density is $N_{\rm H}^{\rm los} = 7.59 \times 10^{23} \,{\rm cm^{-2}}$.
The results of this fit are summarized in Table\,\ref{tab:spectral_fitting}.

Comparing the two models, we can notice that the values of the two \texttt{apec} models and the Gaussian line are consistent within their uncertainties.
First, the inclination angle is slightly smaller in the case of the \rxagn{} model.
For the photon index, the two values lie within the uncertainties.
Yet, the parameters that are related to the circumnuclear medium suggest two different configurations.
In the case of the \rxtopo{} model we find a torus that is Compton thick (i.e., $N_{\rm H}^{\rm tor} > 10^{24}\,{\rm cm^{-2}}$) and has an intermediate covering factor ($CF^{\rm tor} = 0.66^{+0.02}_{-0.01}$).
On the other hand, \rxagn{} returns a torus that is thinner, yet close to the $10^{24}$ limit, with a larger covering factor.
Moreover, the polar cone in the \rxagn{} model is denser than the one in the \rxtopo{} model.
Overall, our results show that for NGC\,424 we have two scenarios, one that suggests a thick torus but in general low total coverage (torus + cone) and another one that has thinner torus but high total coverage, with extra contribution from the BLR and the accretion disk.
The Compton thick scenario of \rxtopo{} is consistent with the literature (\citealt{Matt+:2000BeppoCT,Balokovic+:2014_CTsources,Paltani&Ricci:2017}).
The \rxagn{} model provides a more complex scenario, where the BLR and the accretion disk contribute to the total absorption, and the torus is thinner but with a larger covering factor.
The scope of this analysis is to demonstrate the capabilities of the two new models, and we leave a detailed analysis of NGC\,424 to future studies.

\begin{table}[]
\centering
\caption{Values obtained from the spectral fitting according to the spectral model $tbabs\times(rxtopo/rxagn + apec + apec + zgauss)$.}
\begin{tabular}{lllc}
\hline
\hline
 & \textbf{Model}           & \textbf{RXToPo}  & \textbf{RXagn1}    \\
\hline
\footnotesize{(1)} & Angle $(\theta_{\circ})$ & $79.00^{\circ\,+1}_{\;\;-1}$       & $74.9^{\circ\,+0.4}_{\;\;-2.0}$       \\[3pt]
\footnotesize{(2)} & $\Gamma$              & $1.80^{+0.01}_{-0.01}$     & $1.78^{+0.01}_{-0.01}$      \\[3pt]
\footnotesize{(3)} & $\log \left(N_{\rm H}^{\rm tor}/{\rm cm^{-2}}\right)$    & $24.48^{+0.03}_{-0.08}$    & $23.90^{+0.01}_{-0.01}$    \\[3pt]
\footnotesize{(4)} & $CF_{\rm tor}$      & $0.66^{+0.02}_{-0.01}$     & $0.90^{+u}_{-0.01}$   \\[3pt]
\footnotesize{(5)} & $\log \left(N_{\rm H}^{\rm pol}/{\rm cm^{-2}}\right)$     & $23.19^{+0.04}_{-0.06}$    & $23.80^{+0.01}_{-0.01}$  \\[3pt]
\hline
\footnotesize{(6)} & kT (keV) \texttt{apec} \#1     & $0.10^{+0.01}_{-0.01}$     & $0.10^{+0.01}_{-0.01}$     \\[3pt]
\footnotesize{(7)} & kT (keV) \texttt{apec} \#2     & $0.77^{+0.07}_{-0.08}$    & $0.76^{+0.06}_{-0.07}$   \\[3pt]
\hline
\footnotesize{(8)} & Line energy (keV)  & $6.91^{+0.09}_{-0.14}$     & $6.89^{+0.08}_{-0.11}$    \\[3pt]
\hline
\footnotesize{(9)} & $\chi^2 /dof$  & $136.05/126$ & $154.39/126$   \\
\hline
\end{tabular}
\begin{tablenotes}
\small
\item The model parameters are: (1) the observing angle of the system in degrees, (2) the photon index of the cutoff power-law component, (3) the equatorial hydrogen column density of the dusty torus, (4) the covering factor of the torus, (5) the hydrogen column density of the hollow cone, (6-7) the temperature of the two \texttt{apec} components, (8) the energy of the emission line, (9) ratio between $\chi^2$ and $dof$. The index $u$ means that the upper limit of the parameter was unconstrained by XSPEC.  
\end{tablenotes}
\label{tab:spectral_fitting}
\end{table}

\section{Summary}\label{sec:summary}

The analysis of X-ray spectra of heavily obscured AGN often relies on phenomenological models or simplified geometrical models. These models are frequently limited in terms of flexibility and in the range of physical processes they incorporate.
In this work, we develop two new models (\rxtopo{} and \rxagn{}) using the ray-tracing platform \Reflex{} \citep{Paltani&Ricci:2017}. The two models cover a wide range of energies, extending from 0.3 to 300\,keV and consider moderate energy resolution (20\,eV in the 0.3--10\,keV range, and a logarithmic binning of 0.01 above $10$\,keV).
Both models, \rxtopo{} and \rxagn{}, include material that consists of neutral gas and dust featuring all the major interactions between photons and matter (i.e. Compton scattering, Rayleigh scattering, scattering on dust etc; details in \citealt{Ricci&Paltani:2023}).
First, the \rxtopo{} model includes a central X-ray corona, a dusty torus and a hollow polar conical structure as illustrated in Fig.\,\ref{fig:rxtopo_cartoon} (details in \S\,\ref{sec:rxtopo}).
The free parameters of the model are the observing angle ($\theta_{\circ}$), the photon index ($\Gamma$), the equatorial hydrogen column density of the torus [$N_{\rm H}^{\rm tor}$ in $\log(N_{\rm H}^{\rm tor}/{\rm cm^{-2}})$], the covering factor, calculated as the ratio between the inner radius of the toroidal cutout and the distance from the SMBH ($CF^{\rm tor}=r/R$), and the hydrogen column density of the hollow dusty cone along the inner surface [$N_{\rm H}^{\rm pol}$ in $\log(N_{\rm H}^{\rm pol}/{\rm cm^{-2}})$].
All the parameters are presented in detail in Table\,\ref{Tab:rxtopo components}.

\rxagn{} is an extension of the \rxtopo{} model (Fig.\,\ref{fig:rxagn_cartoon}).
It has the same torus and cone components as \rxtopo{} plus an accretion disk and a broad line region, both of which have fixed geometry and physical parameters (\S\,\ref{sec:rxagn} and Table\,\ref{Tab:rxagn_components}).
As a result, the free parameters are the same as those of the \rxtopo{} model.

As described in \S\,\ref{sec:different_components} five different "flavors" are provided based on the type and number of interactions the photons have undergone: all photons (ALL), photons that have undergone at least an interaction with the material (RPRC), photons that have undergone at least one scattering (SCAT) excluding fluorescent photos, the fluorescence event (FLUO) and the transmitted component which consists of photons that did not interact with the material (CNT). This variety of options allows the user to select the one that is most appropriate for the data being analyzed and/or for the physical scenario they wish to adopt.

We tested these new spectral models on NGC\,424 (see \S\,\ref{sec:fitNGC424}).
This obscured AGN is known to be dominated by reflection, and therefore it is a good candidate to benchmark our models.
We found that both models can fit well the combined XMM-\textit{Newton} and \textit{NuSTAR} data as illustrated in Fig.\,\ref{fig:rxtopo_fit} and \ref{fig:rxagn_fit}.

Our newly developed models have the potential to constrain the geometry and properties of the circumnuclear material around SMBHs by considering very realistic geometries and physical processes. These models can be used to take advantage of broadband X-ray spectroscopy and, in the near future, high-resolution X-ray observations, such as those that will be carried out by the newly launched \textit{XRISM} observatory and future missions like \textit{Athena}.

\begin{acknowledgements}
GD acknowledges support from ANID Beca Doctorado Nacional 21211606.
The research leading to these results has received funding from the European Union's Horizon 2020 Programme under the AHEAD2020 project (grant agreement n. 871158).
CR acknowledges support from Fondecyt Regular grant 1230345, ANID BASAL project FB210003 and the China-Chile joint research fund. 
\end{acknowledgements}

\bibliography{my_references}

@ARTICLE{Gupta+:2021,
       author = {{Gupta}, K.~K. and {Ricci}, C. and {Tortosa}, A. and {Ueda}, Y. and {Kawamuro}, T. and {Koss}, M. and {Trakhtenbrot}, B. and {Oh}, K. and {Bauer}, F.~E. and {Ricci}, F. and {Privon}, G.~C. and {Zappacosta}, L. and {Stern}, D. and {Kakkad}, D. and {Piconcelli}, E. and {Veilleux}, S. and {Mushotzky}, R. and {Caglar}, T. and {Ichikawa}, K. and {Elagali}, A. and {Powell}, M.~C. and {Urry}, C.~M. and {Harrison}, F.},
        title = "{BAT AGN Spectroscopic Survey XXVII: scattered X-Ray radiation in obscured active galactic nuclei}",
      journal = {\mnras},
         year = 2021,
        month = jun,
       volume = {504},
       number = {1},
        pages = {428-443},
          doi = {10.1093/mnras/stab839},
archivePrefix = {arXiv},
       eprint = {2103.10543},
 primaryClass = {astro-ph.HE},
       adsurl = {https://ui.adsabs.harvard.edu/abs/2021MNRAS.504..428G}
}

@ARTICLE{Haardt&Maraschi:1991,
       author = {{Haardt}, F. and {Maraschi}, L.},
        title = "{A Two-Phase Model for the X-Ray Emission from Seyfert Galaxies}",
      journal = {\apjl},
         year = 1991,
        month = oct,
       volume = {380},
        pages = {L51},
          doi = {10.1086/186171},
       adsurl = {https://ui.adsabs.harvard.edu/abs/1991ApJ...380L..51H}
}

@INPROCEEDINGS{XSPEC,
       author = {{Arnaud}, K.~A.},
        title = "{XSPEC: The First Ten Years}",
    booktitle = {Astronomical Data Analysis Software and Systems V},
         year = 1996,
       editor = {{Jacoby}, George H. and {Barnes}, Jeannette},
       series = {Astronomical Society of the Pacific Conference Series},
       volume = {101},
        month = jan,
        pages = {17},
       adsurl = {https://ui.adsabs.harvard.edu/abs/1996ASPC..101...17A}
}

@ARTICLE{Stalevski+:2019,
       author = {{Stalevski}, Marko and {Tristram}, Konrad R.~W. and {Asmus}, Daniel},
        title = "{Dissecting the active galactic nucleus in Circinus - II. A thin dusty disc and a polar outflow on parsec scales}",
      journal = {\mnras},
         year = 2019,
        month = apr,
       volume = {484},
       number = {3},
        pages = {3334-3355},
          doi = {10.1093/mnras/stz220},
archivePrefix = {arXiv},
       eprint = {1901.05488},
 primaryClass = {astro-ph.GA},
       adsurl = {https://ui.adsabs.harvard.edu/abs/2019MNRAS.484.3334S}
}

@ARTICLE{Magdziarz&Zdziarski:1995,
       author = {{Magdziarz}, Pawel and {Zdziarski}, Andrzej A.},
        title = "{Angle-dependent Compton reflection of X-rays and gamma-rays}",
      journal = {\mnras},
         year = 1995,
        month = apr,
       volume = {273},
       number = {3},
        pages = {837-848},
          doi = {10.1093/mnras/273.3.837},
       adsurl = {https://ui.adsabs.harvard.edu/abs/1995MNRAS.273..837M}
}

@ARTICLE{Stalevski+:2017,
       author = {{Stalevski}, Marko and {Asmus}, Daniel and {Tristram}, Konrad R.~W.},
        title = "{Dissecting the active galactic nucleus in Circinus - I. Peculiar mid-IR morphology explained by a dusty hollow cone}",
      journal = {\mnras},
         year = 2017,
        month = dec,
       volume = {472},
       number = {4},
        pages = {3854-3870},
          doi = {10.1093/mnras/stx2227},
archivePrefix = {arXiv},
       eprint = {1708.07838},
 primaryClass = {astro-ph.GA},
       adsurl = {https://ui.adsabs.harvard.edu/abs/2017MNRAS.472.3854S}
}

@ARTICLE{Wada+:2009,
       author = {{Wada}, Keiichi and {Papadopoulos}, Padeli P. and {Spaans}, Marco},
        title = "{Molecular Gas Disk Structures Around Active Galactic Nuclei}",
      journal = {\apj},
         year = 2009,
        month = sep,
       volume = {702},
       number = {1},
        pages = {63-74},
          doi = {10.1088/0004-637X/702/1/63},
archivePrefix = {arXiv},
       eprint = {0906.5444},
 primaryClass = {astro-ph.GA},
       adsurl = {https://ui.adsabs.harvard.edu/abs/2009ApJ...702...63W}
}

@ARTICLE{Ricci+:2015,
       author = {{Ricci}, C. and {Ueda}, Y. and {Koss}, M.~J. and {Trakhtenbrot}, B. and {Bauer}, F.~E. and {Gandhi}, P.},
        title = "{Compton-thick Accretion in the Local Universe}",
      journal = {\apjl},
         year = 2015,
        month = dec,
       volume = {815},
       number = {1},
          eid = {L13},
        pages = {L13},
          doi = {10.1088/2041-8205/815/1/L13},
archivePrefix = {arXiv},
       eprint = {1603.04852},
 primaryClass = {astro-ph.HE},
       adsurl = {https://ui.adsabs.harvard.edu/abs/2015ApJ...815L..13R}
}

@ARTICLE{Goad+:2012,
       author = {{Goad}, M.~R. and {Korista}, K.~T. and {Ruff}, A.~J.},
        title = "{The broad emission-line region: the confluence of the outer accretion disc with the inner edge of the dusty torus}",
      journal = {\mnras},
         year = 2012,
        month = nov,
       volume = {426},
       number = {4},
        pages = {3086-3111},
          doi = {10.1111/j.1365-2966.2012.21808.x},
archivePrefix = {arXiv},
       eprint = {1207.6339},
 primaryClass = {astro-ph.CO},
       adsurl = {https://ui.adsabs.harvard.edu/abs/2012MNRAS.426.3086G}
}

@ARTICLE{Ferland+:1992,
       author = {{Ferland}, G.~J. and {Peterson}, B.~M. and {Horne}, K. and {Welsh}, W.~F. and {Nahar}, S.~N.},
        title = "{Anisotropic Line Emission and the Geometry of the Broad-Line Region in Active Galactic Nuclei}",
      journal = {\apj},
         year = 1992,
        month = mar,
       volume = {387},
        pages = {95},
          doi = {10.1086/171063},
       adsurl = {https://ui.adsabs.harvard.edu/abs/1992ApJ...387...95F}
}

@ARTICLE{Garcia+:2013,
       author = {{Garc{\'\i}a}, J. and {Dauser}, T. and {Reynolds}, C.~S. and {Kallman}, T.~R. and {McClintock}, J.~E. and {Wilms}, J. and {Eikmann}, W.},
        title = "{X-Ray Reflected Spectra from Accretion Disk Models. III. A Complete Grid of Ionized Reflection Calculations}",
      journal = {\apj},
         year = 2013,
        month = may,
       volume = {768},
       number = {2},
          eid = {146},
        pages = {146},
          doi = {10.1088/0004-637X/768/2/146},
archivePrefix = {arXiv},
       eprint = {1303.2112},
 primaryClass = {astro-ph.HE},
       adsurl = {https://ui.adsabs.harvard.edu/abs/2013ApJ...768..146G}
}

@ARTICLE{Koss+:2017,
       author = {{Koss}, Michael and {Trakhtenbrot}, Benny and {Ricci}, Claudio and {Lamperti}, Isabella and {Oh}, Kyuseok and {Berney}, Simon and {Schawinski}, Kevin and {Balokovi{\'c}}, Mislav and {Baronchelli}, Linda and {Crenshaw}, D. Michael and {Fischer}, Travis and {Gehrels}, Neil and {Harrison}, Fiona and {Hashimoto}, Yasuhiro and {Hogg}, Drew and {Ichikawa}, Kohei and {Masetti}, Nicola and {Mushotzky}, Richard and {Sartori}, Lia and {Stern}, Daniel and {Treister}, Ezequiel and {Ueda}, Yoshihiro and {Veilleux}, Sylvain and {Winter}, Lisa},
        title = "{BAT AGN Spectroscopic Survey. I. Spectral Measurements, Derived Quantities, and AGN Demographics}",
      journal = {\apj},
         year = 2017,
        month = nov,
       volume = {850},
       number = {1},
          eid = {74},
        pages = {74},
          doi = {10.3847/1538-4357/aa8ec9},
archivePrefix = {arXiv},
       eprint = {1707.08123},
 primaryClass = {astro-ph.HE},
       adsurl = {https://ui.adsabs.harvard.edu/abs/2017ApJ...850...74K}
}

@ARTICLE{Davies+:2015,
       author = {{Davies}, R.~I. and {Burtscher}, L. and {Rosario}, D. and {Storchi-Bergmann}, T. and {Contursi}, A. and {Genzel}, R. and {Graci{\'a}-Carpio}, J. and {Hicks}, E. and {Janssen}, A. and {Koss}, M. and {Lin}, M. -Y. and {Lutz}, D. and {Maciejewski}, W. and {M{\"u}ller-S{\'a}nchez}, F. and {Orban de Xivry}, G. and {Ricci}, C. and {Riffel}, R. and {Riffel}, R.~A. and {Schartmann}, M. and {Schnorr-M{\"u}ller}, A. and {Sternberg}, A. and {Sturm}, E. and {Tacconi}, L. and {Veilleux}, S.},
        title = "{Insights on the Dusty Torus and Neutral Torus from Optical and X-Ray Obscuration in a Complete Volume Limited Hard X-Ray AGN Sample}",
      journal = {\apj},
         year = 2015,
        month = jun,
       volume = {806},
       number = {1},
          eid = {127},
        pages = {127},
          doi = {10.1088/0004-637X/806/1/127},
archivePrefix = {arXiv},
       eprint = {1505.00536},
 primaryClass = {astro-ph.GA},
       adsurl = {https://ui.adsabs.harvard.edu/abs/2015ApJ...806..127D}
}

@ARTICLE{Arevalo+:2014,
       author = {{Ar{\'e}valo}, P. and {Bauer}, F.~E. and {Puccetti}, S. and {Walton}, D.~J. and {Koss}, M. and {Boggs}, S.~E. and {Brandt}, W.~N. and {Brightman}, M. and {Christensen}, F.~E. and {Comastri}, A. and {Craig}, W.~W. and {Fuerst}, F. and {Gandhi}, P. and {Grefenstette}, B.~W. and {Hailey}, C.~J. and {Harrison}, F.~A. and {Luo}, B. and {Madejski}, G. and {Madsen}, K.~K. and {Marinucci}, A. and {Matt}, G. and {Saez}, C. and {Stern}, D. and {Stuhlinger}, M. and {Treister}, E. and {Urry}, C.~M. and {Zhang}, W.~W.},
        title = "{The 2-79 keV X-Ray Spectrum of the Circinus Galaxy with NuSTAR, XMM-Newton, and Chandra: A Fully Compton-thick Active Galactic Nucleus}",
      journal = {\apj},
         year = 2014,
        month = aug,
       volume = {791},
       number = {2},
          eid = {81},
        pages = {81},
          doi = {10.1088/0004-637X/791/2/81},
archivePrefix = {arXiv},
       eprint = {1406.3345},
 primaryClass = {astro-ph.HE},
       adsurl = {https://ui.adsabs.harvard.edu/abs/2014ApJ...791...81A}
}

@ARTICLE{uttley+:2014,
       author = {{Uttley}, P. and {Cackett}, E.~M. and {Fabian}, A.~C. and {Kara}, E. and {Wilkins}, D.~R.},
        title = "{X-ray reverberation around accreting black holes}",
      journal = {\aapr},
         year = 2014,
        month = aug,
       volume = {22},
          eid = {72},
        pages = {72},
          doi = {10.1007/s00159-014-0072-0},
archivePrefix = {arXiv},
       eprint = {1405.6575},
 primaryClass = {astro-ph.HE},
       adsurl = {https://ui.adsabs.harvard.edu/abs/2014A&ARv..22...72U}
}

@ARTICLE{emmanoulopoulos+:2014,
       author = {{Emmanoulopoulos}, D. and {Papadakis}, I.~E. and {Dov{\v{c}}iak}, M. and {McHardy}, I.~M.},
        title = "{General relativistic modelling of the negative reverberation X-ray time delays in AGN}",
      journal = {\mnras},
         year = 2014,
        month = apr,
       volume = {439},
       number = {4},
        pages = {3931-3950},
          doi = {10.1093/mnras/stu249},
archivePrefix = {arXiv},
       eprint = {1402.0899},
 primaryClass = {astro-ph.HE},
       adsurl = {https://ui.adsabs.harvard.edu/abs/2014MNRAS.439.3931E}
}

@article{Antonucci:1993,
    author = {Antonucci, Robert},
    title = {Unified Models for Active Galactic Nuclei and Quasars},
    journal = {Annual Review of Astronomy and Astrophysics},
    volume = {31},
    number = {1},
    pages = {473-521},
    year = {1993},
    doi = {10.1146/annurev.aa.31.090193.002353},
    URL = {https://doi.org/10.1146/annurev.aa.31.090193.002353},
    eprint = {https://doi.org/10.1146/annurev.aa.31.090193.002353}}

@ARTICLE{Shu:2011,
       author = {{Shu}, X.~W. and {Yaqoob}, T. and {Wang}, J.~X.},
        title = "{Chandra High-energy Grating Observations of the Fe K{\ensuremath{\alpha}} Line Core in Type II Seyfert Galaxies: A Comparison with Type I Nuclei}",
      journal = {\apj},
         year = 2011,
        month = sep,
       volume = {738},
       number = {2},
          eid = {147},
        pages = {147},
          doi = {10.1088/0004-637X/738/2/147},
archivePrefix = {arXiv},
       eprint = {1107.0195},
 primaryClass = {astro-ph.CO},
       adsurl = {https://ui.adsabs.harvard.edu/abs/2011ApJ...738..147S}}

@ARTICLE{Piconcelli2005,
       author = {{Piconcelli}, E. and {Jimenez-Bail{\'o}n}, E. and {Guainazzi}, M. and {Schartel}, N. and {Rodr{\'\i}guez-Pascual}, P.~M. and {Santos-Lle{\'o}}, M.},
        title = "{The XMM-Newton view of PG quasars.  I. X-ray continuum and absorption}",
      journal = {\aap},
         year = 2005,
        month = mar,
       volume = {432},
       number = {1},
        pages = {15-30},
          doi = {10.1051/0004-6361:20041621},
archivePrefix = {arXiv},
       eprint = {astro-ph/0411051},
 primaryClass = {astro-ph},
       adsurl = {https://ui.adsabs.harvard.edu/abs/2005A&A...432...15P}
}

@ARTICLE{Laloux:2024,
       author = {{Laloux}, Brivael and {Georgakakis}, Antonis and {Alexander}, David M. and {Buchner}, Johannes and {Andonie}, Carolina and {Acharya}, Nischal and {Aird}, James and {Alonso-Tetilla}, Alba V. and {Bongiorno}, Angela and {Hickox}, Ryan C. and {Lapi}, Andrea and {Musiimenta}, Blessing and {Ramos Almeida}, Cristina and {Vellforth}, Carolin and {Shankar}, Francesco},
        title = "{Accretion properties of X-ray AGN: Evidence for radiation-regulated obscuration with redshift-dependent host galaxy contribution}",
      journal = {arXiv e-prints},
         year = 2024,
        month = mar,
          eid = {arXiv:2403.07109},
        pages = {arXiv:2403.07109},
          doi = {10.48550/arXiv.2403.07109},
archivePrefix = {arXiv},
       eprint = {2403.07109},
 primaryClass = {astro-ph.GA},
       adsurl = {https://ui.adsabs.harvard.edu/abs/2024arXiv240307109L}
}

@ARTICLE{Ricci:2017Nat,
       author = {{Ricci}, Claudio and {Trakhtenbrot}, Benny and {Koss}, Michael J. and {Ueda}, Yoshihiro and {Schawinski}, Kevin and {Oh}, Kyuseok and {Lamperti}, Isabella and {Mushotzky}, Richard and {Treister}, Ezequiel and {Ho}, Luis C. and {Weigel}, Anna and {Bauer}, Franz E. and {Paltani}, Stephane and {Fabian}, Andrew C. and {Xie}, Yanxia and {Gehrels}, Neil},
        title = "{The close environments of accreting massive black holes are shaped by radiative feedback}",
      journal = {\nat},
         year = 2017,
        month = sep,
       volume = {549},
       number = {7673},
        pages = {488-491},
          doi = {10.1038/nature23906},
archivePrefix = {arXiv},
       eprint = {1709.09651},
 primaryClass = {astro-ph.HE},
       adsurl = {https://ui.adsabs.harvard.edu/abs/2017Natur.549..488R}
}

@ARTICLE{Ricci:2022,
       author = {{Ricci}, C. and {Ananna}, T.~T. and {Temple}, M.~J. and {Urry}, C.~M. and {Koss}, M.~J. and {Trakhtenbrot}, B. and {Ueda}, Y. and {Stern}, D. and {Bauer}, F.~E. and {Treister}, E. and {Privon}, G.~C. and {Oh}, K. and {Paltani}, S. and {Stalevski}, M. and {Ho}, L.~C. and {Fabian}, A.~C. and {Mushotzky}, R. and {Chang}, C.~S. and {Ricci}, F. and {Kakkad}, D. and {Sartori}, L. and {Baer}, R. and {Caglar}, T. and {Powell}, M. and {Harrison}, F.},
        title = "{BASS XXXVII: The Role of Radiative Feedback in the Growth and Obscuration Properties of Nearby Supermassive Black Holes}",
      journal = {\apj},
         year = 2022,
        month = oct,
       volume = {938},
       number = {1},
          eid = {67},
        pages = {67},
          doi = {10.3847/1538-4357/ac8e67},
archivePrefix = {arXiv},
       eprint = {2209.00014},
 primaryClass = {astro-ph.GA},
       adsurl = {https://ui.adsabs.harvard.edu/abs/2022ApJ...938...67R}
}

@ARTICLE{Venanzi:2020,
       author = {{Venanzi}, Marta and {H{\"o}nig}, Sebastian and {Williamson}, David},
        title = "{The Role of Infrared Radiation Pressure in Shaping Dusty Winds in AGNs}",
      journal = {\apj},
         year = 2020,
        month = sep,
       volume = {900},
       number = {2},
          eid = {174},
        pages = {174},
          doi = {10.3847/1538-4357/aba89f},
archivePrefix = {arXiv},
       eprint = {2007.13554},
 primaryClass = {astro-ph.GA},
       adsurl = {https://ui.adsabs.harvard.edu/abs/2020ApJ...900..174V}
}

@ARTICLE{Hoenig:2007,
       author = {{H{\"o}nig}, S.~F. and {Beckert}, T.},
        title = "{Active galactic nuclei dust tori at low and high luminosities}",
      journal = {\mnras},
         year = 2007,
        month = sep,
       volume = {380},
       number = {3},
        pages = {1172-1176},
          doi = {10.1111/j.1365-2966.2007.12157.x},
       adsurl = {https://ui.adsabs.harvard.edu/abs/2007MNRAS.380.1172H}
}

@ARTICLE{Fabian:2006,
       author = {{Fabian}, A.~C. and {Celotti}, A. and {Erlund}, M.~C.},
        title = "{Radiative pressure feedback by a quasar in a galactic bulge}",
      journal = {\mnras},
         year = 2006,
        month = nov,
       volume = {373},
       number = {1},
        pages = {L16-L20},
          doi = {10.1111/j.1745-3933.2006.00234.x},
archivePrefix = {arXiv},
       eprint = {astro-ph/0608425},
 primaryClass = {astro-ph},
       adsurl = {https://ui.adsabs.harvard.edu/abs/2006MNRAS.373L..16F}
}

@ARTICLE{Murphy&Yaqoob:2009,
       author = {{Murphy}, Kendrah D. and {Yaqoob}, Tahir},
        title = "{An X-ray spectral model for Compton-thick toroidal reprocessors}",
      journal = {\mnras},
         year = 2009,
        month = aug,
       volume = {397},
       number = {3},
        pages = {1549-1562},
          doi = {10.1111/j.1365-2966.2009.15025.x},
archivePrefix = {arXiv},
       eprint = {0905.3188},
 primaryClass = {astro-ph.HE},
       adsurl = {https://ui.adsabs.harvard.edu/abs/2009MNRAS.397.1549M}}

@ARTICLE{Paltani&Ricci:2017,
       author = {{Paltani}, S. and {Ricci}, C.},
        title = "{RefleX: X-ray absorption and reflection in active galactic nuclei for arbitrary geometries}",
      journal = {\aap},
         year = 2017,
        month = nov,
       volume = {607},
          eid = {A31},
        pages = {A31},
          doi = {10.1051/0004-6361/201629623},
archivePrefix = {arXiv},
       eprint = {1906.08824},
 primaryClass = {astro-ph.HE},
       adsurl = {https://ui.adsabs.harvard.edu/abs/2017A&A...607A..31P}}

@ARTICLE{Chartas:2009,
       author = {{Chartas}, G. and {Kochanek}, C.~S. and {Dai}, X. and {Poindexter}, S. and {Garmire}, G.},
        title = "{X-Ray Microlensing in RXJ1131-1231 and HE1104-1805}",
      journal = {\apj},
         year = 2009,
        month = mar,
       volume = {693},
       number = {1},
        pages = {174-185},
          doi = {10.1088/0004-637X/693/1/174},
archivePrefix = {arXiv},
       eprint = {0805.4492},
 primaryClass = {astro-ph},
       adsurl = {https://ui.adsabs.harvard.edu/abs/2009ApJ...693..174C}}

@ARTICLE{Fabian:2009,
       author = {{Fabian}, A.~C. and {Zoghbi}, A. and {Ross}, R.~R. and {Uttley}, P. and {Gallo}, L.~C. and {Brandt}, W.~N. and {Blustin}, A.~J. and {Boller}, T. and {Caballero-Garcia}, M.~D. and {Larsson}, J. and {Miller}, J.~M. and {Miniutti}, G. and {Ponti}, G. and {Reis}, R.~C. and {Reynolds}, C.~S. and {Tanaka}, Y. and {Young}, A.~J.},
        title = "{Broad line emission from iron K- and L-shell transitions in the active galaxy 1H0707-495}",
      journal = {\nat},
         year = 2009,
        month = may,
       volume = {459},
       number = {7246},
        pages = {540-542},
          doi = {10.1038/nature08007},
       adsurl = {https://ui.adsabs.harvard.edu/abs/2009Natur.459..540F}}

@ARTICLE{DeMarco:2013,
       author = {{De Marco}, B. and {Ponti}, G. and {Cappi}, M. and {Dadina}, M. and {Uttley}, P. and {Cackett}, E.~M. and {Fabian}, A.~C. and {Miniutti}, G.},
        title = "{Discovery of a relation between black hole mass and soft X-ray time lags in active galactic nuclei}",
      journal = {\mnras},
         year = 2013,
        month = may,
       volume = {431},
       number = {3},
        pages = {2441-2452},
          doi = {10.1093/mnras/stt339},
archivePrefix = {arXiv},
       eprint = {1201.0196},
 primaryClass = {astro-ph.HE},
       adsurl = {https://ui.adsabs.harvard.edu/abs/2013MNRAS.431.2441D}}

@Article{Haardt&mar:1993,
  author   = {{Haardt}, F. and {Maraschi}, L.},
  title    = {{X-ray spectra from two-phase accretion disks}},
  journal  = {\apj},
  year     = {1993},
  volume   = {413},
  pages    = {507-517},
  month    = aug,
  adsurl   = {http://adsabs.harvard.edu/abs/1993ApJ...413..507H},
  doi      = {10.1086/173020}}

@ARTICLE{Wilms:2000,
       author = {{Wilms}, J. and {Allen}, A. and {McCray}, R.},
        title = "{On the Absorption of X-Rays in the Interstellar Medium}",
      journal = {\apj},
         year = 2000,
        month = oct,
       volume = {542},
       number = {2},
        pages = {914-924},
          doi = {10.1086/317016},
archivePrefix = {arXiv},
       eprint = {astro-ph/0008425},
 primaryClass = {astro-ph},
       adsurl = {https://ui.adsabs.harvard.edu/abs/2000ApJ...542..914W}}

@ARTICLE{xrism:2020,
       author = {{XRISM Science Team}},
        title = "{Science with the X-ray Imaging and Spectroscopy Mission (XRISM)}",
      journal = {arXiv e-prints},
         year = 2020,
        month = mar,
          eid = {arXiv:2003.04962},
        pages = {arXiv:2003.04962},
archivePrefix = {arXiv},
       eprint = {2003.04962},
 primaryClass = {astro-ph.HE},
       adsurl = {https://ui.adsabs.harvard.edu/abs/2020arXiv200304962X}}

@ARTICLE{Mor:2009,
       author = {{Mor}, Rivay and {Netzer}, Hagai and {Elitzur}, Moshe},
        title = "{Dusty Structure Around Type-I Active Galactic Nuclei: Clumpy Torus Narrow-line Region and Near-nucleus Hot Dust}",
      journal = {\apj},
         year = 2009,
        month = nov,
       volume = {705},
       number = {1},
        pages = {298-313},
          doi = {10.1088/0004-637X/705/1/298},
archivePrefix = {arXiv},
       eprint = {0907.1654},
 primaryClass = {astro-ph.CO},
       adsurl = {https://ui.adsabs.harvard.edu/abs/2009ApJ...705..298M}}

@ARTICLE{Kaspi:2005,
       author = {{Kaspi}, Shai and {Maoz}, Dan and {Netzer}, Hagai and {Peterson}, Bradley M. and {Vestergaard}, Marianne and {Jannuzi}, Buell T.},
        title = "{The Relationship between Luminosity and Broad-Line Region Size in Active Galactic Nuclei}",
      journal = {\apj},
         year = 2005,
        month = aug,
       volume = {629},
       number = {1},
        pages = {61-71},
          doi = {10.1086/431275},
archivePrefix = {arXiv},
       eprint = {astro-ph/0504484},
 primaryClass = {astro-ph},
       adsurl = {https://ui.adsabs.harvard.edu/abs/2005ApJ...629...61K}}

@ARTICLE{Vasudevan:2009,
       author = {{Vasudevan}, R.~V. and {Fabian}, A.~C.},
        title = "{Simultaneous X-ray/optical/UV snapshots of active galactic nuclei from XMM-Newton: spectral energy distributions for the reverberation mapped sample}",
      journal = {\mnras},
         year = 2009,
        month = jan,
       volume = {392},
       number = {3},
        pages = {1124-1140},
          doi = {10.1111/j.1365-2966.2008.14108.x},
archivePrefix = {arXiv},
       eprint = {0810.3777},
 primaryClass = {astro-ph},
       adsurl = {https://ui.adsabs.harvard.edu/abs/2009MNRAS.392.1124V}}

@ARTICLE{Ricci+:2018,
       author = {{Ricci}, C. and {Ho}, L.~C. and {Fabian}, A.~C. and {Trakhtenbrot}, B. and {Koss}, M.~J. and {Ueda}, Y. and {Lohfink}, A. and {Shimizu}, T. and {Bauer}, F.~E. and {Mushotzky}, R. and {Schawinski}, K. and {Paltani}, S. and {Lamperti}, I. and {Treister}, E. and {Oh}, K.},
        title = "{BAT AGN Spectroscopic Survey - XII. The relation between coronal properties of active galactic nuclei and the Eddington ratio}",
      journal = {\mnras},
         year = 2018,
        month = oct,
       volume = {480},
       number = {2},
        pages = {1819-1830},
          doi = {10.1093/mnras/sty1879},
archivePrefix = {arXiv},
       eprint = {1809.04076},
 primaryClass = {astro-ph.HE},
       adsurl = {https://ui.adsabs.harvard.edu/abs/2018MNRAS.480.1819R}
}

@ARTICLE{Urry&Padovani:1995,
       author = {{Urry}, C. Megan and {Padovani}, Paolo},
        title = "{Unified Schemes for Radio-Loud Active Galactic Nuclei}",
      journal = {\pasp},
         year = 1995,
        month = sep,
       volume = {107},
        pages = {803},
          doi = {10.1086/133630},
archivePrefix = {arXiv},
       eprint = {astro-ph/9506063},
 primaryClass = {astro-ph},
       adsurl = {https://ui.adsabs.harvard.edu/abs/1995PASP..107..803U}
}

@ARTICLE{RamosAlmeida&Ricci:2017Nat,
       author = {{Ramos Almeida}, Cristina and {Ricci}, Claudio},
        title = "{Nuclear obscuration in active galactic nuclei}",
      journal = {Nature Astronomy},
         year = 2017,
        month = oct,
       volume = {1},
        pages = {679-689},
          doi = {10.1038/s41550-017-0232-z},
archivePrefix = {arXiv},
       eprint = {1709.00019},
 primaryClass = {astro-ph.GA},
       adsurl = {https://ui.adsabs.harvard.edu/abs/2017NatAs...1..679R}
}

@ARTICLE{GravityCollab:2018Natur,
       author = {{Gravity Collaboration} and {Sturm}, E. and {Dexter}, J. and {Pfuhl}, O. and {Stock}, M.~R. and {Davies}, R.~I. and {Lutz}, D. and {Cl{\'e}net}, Y. and {Eckart}, A. and {Eisenhauer}, F. and {Genzel}, R. and {Gratadour}, D. and {H{\"o}nig}, S.~F. and {Kishimoto}, M. and {Lacour}, S. and {Millour}, F. and {Netzer}, H. and {Perrin}, G. and {Peterson}, B.~M. and {Petrucci}, P.~O. and {Rouan}, D. and {Waisberg}, I. and {Woillez}, J. and {Amorim}, A. and {Brandner}, W. and {F{\"o}rster Schreiber}, N.~M. and {Garcia}, P.~J.~V. and {Gillessen}, S. and {Ott}, T. and {Paumard}, T. and {Perraut}, K. and {Scheithauer}, S. and {Straubmeier}, C. and {Tacconi}, L.~J. and {Widmann}, F.},
        title = "{Spatially resolved rotation of the broad-line region of a quasar at sub-parsec scale}",
      journal = {\nat},
         year = 2018,
        month = nov,
       volume = {563},
       number = {7733},
        pages = {657-660},
          doi = {10.1038/s41586-018-0731-9},
archivePrefix = {arXiv},
       eprint = {1811.11195},
 primaryClass = {astro-ph.GA},
       adsurl = {https://ui.adsabs.harvard.edu/abs/2018Natur.563..657G}
}

@ARTICLE{GravityColab:2020,
       author = {{GRAVITY Collaboration} and {Amorim}, A. and {Baub{\"o}ck}, M. and {Brandner}, W. and {Cl{\'e}net}, Y. and {Davies}, R. and {de Zeeuw}, P.~T. and {Dexter}, J. and {Eckart}, A. and {Eisenhauer}, F. and {F{\"o}rster Schreiber}, N.~M. and {Gao}, F. and {Garcia}, P.~J.~V. and {Genzel}, R. and {Gillessen}, S. and {Gratadour}, D. and {H{\"o}nig}, S. and {Kishimoto}, M. and {Lacour}, S. and {Lutz}, D. and {Millour}, F. and {Netzer}, H. and {Ott}, T. and {Paumard}, T. and {Perraut}, K. and {Perrin}, G. and {Peterson}, B.~M. and {Petrucci}, P.~O. and {Pfuhl}, O. and {Prieto}, M.~A. and {Rouan}, D. and {Shangguan}, J. and {Shimizu}, T. and {Schartmann}, M. and {Stadler}, J. and {Sternberg}, A. and {Straub}, O. and {Straubmeier}, C. and {Sturm}, E. and {Tacconi}, L.~J. and {Tristram}, K.~R.~W. and {Vermot}, P. and {von Fellenberg}, S. and {Waisberg}, I. and {Widmann}, F. and {Woillez}, J.},
        title = "{The spatially resolved broad line region of IRAS 09149-6206}",
      journal = {\aap},
         year = 2020,
        month = nov,
       volume = {643},
          eid = {A154},
        pages = {A154},
          doi = {10.1051/0004-6361/202039067},
archivePrefix = {arXiv},
       eprint = {2009.08463},
 primaryClass = {astro-ph.GA},
       adsurl = {https://ui.adsabs.harvard.edu/abs/2020A&A...643A.154G}
}

@ARTICLE{Asmus:2019,
       author = {{Asmus}, D.},
        title = "{New evidence for the ubiquity of prominent polar dust emission in AGN on tens of parsec scales}",
      journal = {\mnras},
         year = 2019,
        month = oct,
       volume = {489},
       number = {2},
        pages = {2177-2188},
          doi = {10.1093/mnras/stz2289},
archivePrefix = {arXiv},
       eprint = {1908.03552},
 primaryClass = {astro-ph.GA},
       adsurl = {https://ui.adsabs.harvard.edu/abs/2019MNRAS.489.2177A}
}

@ARTICLE{lpgs,
       author = {{Lodders}, K. and {Palme}, H. and {Gail}, H. -P.},
        title = "{Abundances of the Elements in the Solar System}",
      journal = {Landolt B{\"o}rnstein},
         year = 2009,
        month = jan,
       volume = {4B},
        pages = {712},
          doi = {10.1007/978-3-540-88055-4_34},
archivePrefix = {arXiv},
       eprint = {0901.1149},
 primaryClass = {astro-ph.EP},
       adsurl = {https://ui.adsabs.harvard.edu/abs/2009LanB...4B..712L}
}

@ARTICLE{MONACO:2011,
       author = {{Odaka}, Hirokazu and {Aharonian}, Felix and {Watanabe}, Shin and {Tanaka}, Yasuyuki and {Khangulyan}, Dmitry and {Takahashi}, Tadayuki},
        title = "{X-Ray Diagnostics of Giant Molecular Clouds in the Galactic Center Region and Past Activity of Sgr A*}",
      journal = {\apj},
         year = 2011,
        month = oct,
       volume = {740},
       number = {2},
          eid = {103},
        pages = {103},
          doi = {10.1088/0004-637X/740/2/103},
archivePrefix = {arXiv},
       eprint = {1110.1936},
 primaryClass = {astro-ph.GA},
       adsurl = {https://ui.adsabs.harvard.edu/abs/2011ApJ...740..103O}
}

@article{skirt_I:2003,
    author = {Baes, Maarten and Davies, Jonathan I. and Dejonghe, Herwig and Sabatini, Sabina and Roberts, Sarah and Evans, Rhodri and Linder, Suzanne M. and Smith, Rodney M. and de Blok, W. J. G.},
    title = "{Radiative transfer in disc galaxies — III. The observed kinematics of dusty disc galaxies}",
    journal = {Monthly Notices of the Royal Astronomical Society},
    volume = {343},
    number = {4},
    pages = {1081-1094},
    year = {2003},
    month = {08},
    issn = {0035-8711},
    doi = {10.1046/j.1365-8711.2003.06770.x},
    url = {https://doi.org/10.1046/j.1365-8711.2003.06770.x},
    eprint = {https://academic.oup.com/mnras/article-pdf/343/4/1081/3655000/343-4-1081.pdf},
}

@ARTICLE{skirt_II:2011ApJS,
       author = {{Baes}, Maarten and {Verstappen}, Joris and {De Looze}, Ilse and {Fritz}, Jacopo and {Saftly}, Waad and {Vidal P{\'e}rez}, Edgardo and {Stalevski}, Marko and {Valcke}, Sander},
        title = "{Efficient Three-dimensional NLTE Dust Radiative Transfer with SKIRT}",
      journal = {\apjs},
         year = 2011,
        month = oct,
       volume = {196},
       number = {2},
          eid = {22},
        pages = {22},
          doi = {10.1088/0067-0049/196/2/22},
archivePrefix = {arXiv},
       eprint = {1108.5056},
 primaryClass = {astro-ph.CO},
       adsurl = {https://ui.adsabs.harvard.edu/abs/2011ApJS..196...22B}
}

@ARTICLE{XARS_Buchner:2019A&A,
       author = {{Buchner}, Johannes and {Brightman}, Murray and {Nandra}, Kirpal and {Nikutta}, Robert and {Bauer}, Franz E.},
        title = "{X-ray spectral and eclipsing model of the clumpy obscurer in active galactic nuclei}",
      journal = {\aap},
         year = 2019,
        month = sep,
       volume = {629},
          eid = {A16},
        pages = {A16},
          doi = {10.1051/0004-6361/201834771},
archivePrefix = {arXiv},
       eprint = {1907.13137},
 primaryClass = {astro-ph.HE},
       adsurl = {https://ui.adsabs.harvard.edu/abs/2019A&A...629A..16B}
}

@ARTICLE{SKIRT_Xrays:2023,
       author = {{Vander Meulen}, Bert and {Camps}, Peter and {Stalevski}, Marko and {Baes}, Maarten},
        title = "{X-ray radiative transfer in full 3D with SKIRT}",
      journal = {\aap},
         year = 2023,
        month = jun,
       volume = {674},
          eid = {A123},
        pages = {A123},
          doi = {10.1051/0004-6361/202245783},
archivePrefix = {arXiv},
       eprint = {2304.10563},
 primaryClass = {astro-ph.HE},
       adsurl = {https://ui.adsabs.harvard.edu/abs/2023A&A...674A.123V}
}

@ARTICLE{Zdziarski96,
       author = {{Zdziarski}, A.~A. and {Johnson}, W.~N. and {Magdziarz}, P.},
        title = "{Broad-band {\ensuremath{\gamma}}-ray and X-ray spectra of NGC 4151 and their implications for physical processes and geometry.}",
      journal = {\mnras},
         year = 1996,
        month = nov,
       volume = {283},
       number = {1},
        pages = {193-206},
          doi = {10.1093/mnras/283.1.193},
archivePrefix = {arXiv},
       eprint = {astro-ph/9607015},
 primaryClass = {astro-ph},
       adsurl = {https://ui.adsabs.harvard.edu/abs/1996MNRAS.283..193Z}
}

@ARTICLE{Dadina2008,
       author = {{Dadina}, M.},
        title = "{Seyfert galaxies in the local Universe (z {\ensuremath{\leq}} 0.1): the average X-ray spectrum as seen by BeppoSAX}",
      journal = {\aap},
         year = 2008,
        month = jul,
       volume = {485},
       number = {2},
        pages = {417-424},
          doi = {10.1051/0004-6361:20077569},
archivePrefix = {arXiv},
       eprint = {0801.4338},
 primaryClass = {astro-ph},
       adsurl = {https://ui.adsabs.harvard.edu/abs/2008A&A...485..417D}
}

@ARTICLE{Ricci&Paltani:2023,
       author = {{Ricci}, Claudio and {Paltani}, St{\'e}phane},
        title = "{Ray-tracing Simulations and Spectral Models of X-Ray Radiation in Dusty Media}",
      journal = {\apj},
         year = 2023,
        month = mar,
       volume = {945},
       number = {1},
          eid = {55},
        pages = {55},
          doi = {10.3847/1538-4357/acb5a6},
archivePrefix = {arXiv},
       eprint = {2301.10268},
 primaryClass = {astro-ph.HE},
       adsurl = {https://ui.adsabs.harvard.edu/abs/2023ApJ...945...55R}
}

@ARTICLE{Shakura&Sunyaev:1973,
       author = {{Shakura}, N.~I. and {Sunyaev}, R.~A.},
        title = "{Black holes in binary systems. Observational appearance.}",
      journal = {\aap},
         year = 1973,
        month = jan,
       volume = {24},
        pages = {337-355},
       adsurl = {https://ui.adsabs.harvard.edu/abs/1973A&A....24..337S}
}

@article{Tanimoto:2019,
doi = {10.3847/1538-4357/ab1b20},
url = {https://dx.doi.org/10.3847/1538-4357/ab1b20},
year = {2019},
month = {may},
publisher = {The American Astronomical Society},
volume = {877},
number = {2},
pages = {95},
author = {Atsushi Tanimoto and Yoshihiro Ueda and Hirokazu Odaka and Toshihiro Kawaguchi and Yasushi Fukazawa and Taiki Kawamuro},
title = {XCLUMPY: X-Ray Spectral Model from Clumpy Torus and Its Application to the Circinus Galaxy},
journal = {The Astrophysical Journal},
}

@ARTICLE{Nenkova+:2008,
       author = {{Nenkova}, Maia and {Sirocky}, Matthew M. and {Ivezi{\'c}}, {\v{Z}}eljko and {Elitzur}, Moshe},
        title = "{AGN Dusty Tori. I. Handling of Clumpy Media}",
      journal = {\apj},
         year = 2008,
        month = sep,
       volume = {685},
       number = {1},
        pages = {147-159},
          doi = {10.1086/590482},
archivePrefix = {arXiv},
       eprint = {0806.0511},
 primaryClass = {astro-ph},
       adsurl = {https://ui.adsabs.harvard.edu/abs/2008ApJ...685..147N}
}

@ARTICLE{Balokovic+:2018,
       author = {{Balokovi{\'c}}, M. and {Brightman}, M. and {Harrison}, F.~A. and {Comastri}, A. and {Ricci}, C. and {Buchner}, J. and {Gandhi}, P. and {Farrah}, D. and {Stern}, D.},
        title = "{New Spectral Model for Constraining Torus Covering Factors from Broadband X-Ray Spectra of Active Galactic Nuclei}",
      journal = {\apj},
         year = 2018,
        month = feb,
       volume = {854},
       number = {1},
          eid = {42},
        pages = {42},
          doi = {10.3847/1538-4357/aaa7eb},
archivePrefix = {arXiv},
       eprint = {1801.04938},
 primaryClass = {astro-ph.HE},
       adsurl = {https://ui.adsabs.harvard.edu/abs/2018ApJ...854...42B}
}

@ARTICLE{lpgs:2009,
       author = {{Lodders}, K. and {Palme}, H. and {Gail}, H. -P.},
        title = "{Abundances of the Elements in the Solar System}",
      journal = {Landolt B\&ouml;rnstein},
         year = 2009,
        month = jan,
       volume = {4B},
        pages = {712},
          doi = {10.1007/978-3-540-88055-4_34},
archivePrefix = {arXiv},
       eprint = {0901.1149},
 primaryClass = {astro-ph.EP},
       adsurl = {https://ui.adsabs.harvard.edu/abs/2009LanB...4B..712L}
}

@ARTICLE{McHardy+:2023,
       author = {{McHardy}, I.~M. and {Beard}, M. and {Breedt}, E. and {Knapen}, J.~H. and {Vincentelli}, F.~M. and {Veresvarska}, M. and {Dhillon}, V.~S. and {Marsh}, T.~R. and {Littlefair}, S.~P. and {Horne}, K. and {Glew}, R. and {Goad}, M.~R. and {Kammoun}, E. and {Emmanoulopoulos}, D.},
        title = "{First detection of the outer edge of an AGN accretion disc: very fast multiband optical variability of NGC 4395 with GTC/HiPERCAM and LT/IO:O}",
      journal = {\mnras},
         year = 2023,
        month = mar,
       volume = {519},
       number = {3},
        pages = {3366-3382},
          doi = {10.1093/mnras/stac3651},
archivePrefix = {arXiv},
       eprint = {2212.08015},
 primaryClass = {astro-ph.GA},
       adsurl = {https://ui.adsabs.harvard.edu/abs/2023MNRAS.519.3366M}
}

@ARTICLE{Garcia+:2010,
       author = {{Garc{\'\i}a}, J. and {Kallman}, T.~R.},
        title = "{X-ray Reflected Spectra from Accretion Disk Models. I. Constant Density Atmospheres}",
      journal = {\apj},
         year = 2010,
        month = aug,
       volume = {718},
       number = {2},
        pages = {695-706},
          doi = {10.1088/0004-637X/718/2/695},
archivePrefix = {arXiv},
       eprint = {1006.0485},
 primaryClass = {astro-ph.HE},
       adsurl = {https://ui.adsabs.harvard.edu/abs/2010ApJ...718..695G}
}

@ARTICLE{Garcia+:2014,
       author = {{Garc{\'\i}a}, J. and {Dauser}, T. and {Lohfink}, A. and {Kallman}, T.~R. and {Steiner}, J.~F. and {McClintock}, J.~E. and {Brenneman}, L. and {Wilms}, J. and {Eikmann}, W. and {Reynolds}, C.~S. and {Tombesi}, F.},
        title = "{Improved Reflection Models of Black Hole Accretion Disks: Treating the Angular Distribution of X-Rays}",
      journal = {\apj},
         year = 2014,
        month = feb,
       volume = {782},
       number = {2},
          eid = {76},
        pages = {76},
          doi = {10.1088/0004-637X/782/2/76},
archivePrefix = {arXiv},
       eprint = {1312.3231},
 primaryClass = {astro-ph.HE},
       adsurl = {https://ui.adsabs.harvard.edu/abs/2014ApJ...782...76G}
}

@ARTICLE{Ricci+17Catalog,
       author = {{Ricci}, C. and {Trakhtenbrot}, B. and {Koss}, M.~J. and {Ueda}, Y. and {Del Vecchio}, I. and {Treister}, E. and {Schawinski}, K. and {Paltani}, S. and {Oh}, K. and {Lamperti}, I. and {Berney}, S. and {Gandhi}, P. and {Ichikawa}, K. and {Bauer}, F.~E. and {Ho}, L.~C. and {Asmus}, D. and {Beckmann}, V. and {Soldi}, S. and {Balokovi{\'c}}, M. and {Gehrels}, N. and {Markwardt}, C.~B.},
        title = "{BAT AGN Spectroscopic Survey. V. X-Ray Properties of the Swift/BAT 70-month AGN Catalog}",
      journal = {\apjs},
         year = 2017,
        month = dec,
       volume = {233},
       number = {2},
          eid = {17},
        pages = {17},
          doi = {10.3847/1538-4365/aa96ad},
archivePrefix = {arXiv},
       eprint = {1709.03989},
 primaryClass = {astro-ph.HE},
       adsurl = {https://ui.adsabs.harvard.edu/abs/2017ApJS..233...17R}
}

@ARTICLE{Matt+:2000BeppoCT,
       author = {{Matt}, G. and {Fabian}, A.~C. and {Guainazzi}, M. and {Iwasawa}, K. and {Bassani}, L. and {Malaguti}, G.},
        title = "{The X-ray spectra of Compton-thick Seyfert 2 galaxies as seen by BeppoSAX}",
      journal = {\mnras},
         year = 2000,
        month = oct,
       volume = {318},
       number = {1},
        pages = {173-179},
          doi = {10.1046/j.1365-8711.2000.03721.x},
archivePrefix = {arXiv},
       eprint = {astro-ph/0005219},
 primaryClass = {astro-ph},
       adsurl = {https://ui.adsabs.harvard.edu/abs/2000MNRAS.318..173M}
}

@ARTICLE{Marinucci+:2011ngc424,
       author = {{Marinucci}, A. and {Bianchi}, S. and {Matt}, G. and {Fabian}, A.~C. and {Iwasawa}, K. and {Miniutti}, G. and {Piconcelli}, E.},
        title = "{The X-ray spectral signatures from the complex circumnuclear regions in the Compton thick AGN NGC 424}",
      journal = {\aap},
         year = 2011,
        month = feb,
       volume = {526},
          eid = {A36},
        pages = {A36},
          doi = {10.1051/0004-6361/201015358},
archivePrefix = {arXiv},
       eprint = {1010.4227},
 primaryClass = {astro-ph.CO},
       adsurl = {https://ui.adsabs.harvard.edu/abs/2011A&A...526A..36M}
}

@ARTICLE{Balokovic+:2014_CTsources,
       author = {{Balokovi{\'c}}, M. and {Comastri}, A. and {Harrison}, F.~A. and {Alexander}, D.~M. and {Ballantyne}, D.~R. and {Bauer}, F.~E. and {Boggs}, S.~E. and {Brandt}, W.~N. and {Brightman}, M. and {Christensen}, F.~E. and {Craig}, W.~W. and {Del Moro}, A. and {Gandhi}, P. and {Hailey}, C.~J. and {Koss}, M. and {Lansbury}, G.~B. and {Luo}, B. and {Madejski}, G.~M. and {Marinucci}, A. and {Matt}, G. and {Markwardt}, C.~B. and {Puccetti}, S. and {Reynolds}, C.~S. and {Risaliti}, G. and {Rivers}, E. and {Stern}, D. and {Walton}, D.~J. and {Zhang}, W.~W.},
        title = "{The NuSTAR View of Nearby Compton-thick Active Galactic Nuclei: The Cases of NGC 424, NGC 1320, and IC 2560}",
      journal = {\apj},
         year = 2014,
        month = oct,
       volume = {794},
       number = {2},
          eid = {111},
        pages = {111},
          doi = {10.1088/0004-637X/794/2/111},
archivePrefix = {arXiv},
       eprint = {1408.5414},
 primaryClass = {astro-ph.HE},
       adsurl = {https://ui.adsabs.harvard.edu/abs/2014ApJ...794..111B}
}

@ARTICLE{HI4PI:NH:2016A&A,
       author = {{HI4PI Collaboration} and {Ben Bekhti}, N. and {Fl{\"o}er}, L. and {Keller}, R. and {Kerp}, J. and {Lenz}, D. and {Winkel}, B. and {Bailin}, J. and {Calabretta}, M.~R. and {Dedes}, L. and {Ford}, H.~A. and {Gibson}, B.~K. and {Haud}, U. and {Janowiecki}, S. and {Kalberla}, P.~M.~W. and {Lockman}, F.~J. and {McClure-Griffiths}, N.~M. and {Murphy}, T. and {Nakanishi}, H. and {Pisano}, D.~J. and {Staveley-Smith}, L.},
        title = "{HI4PI: A full-sky H I survey based on EBHIS and GASS}",
      journal = {\aap},
         year = 2016,
        month = oct,
       volume = {594},
          eid = {A116},
        pages = {A116},
          doi = {10.1051/0004-6361/201629178},
archivePrefix = {arXiv},
       eprint = {1610.06175},
 primaryClass = {astro-ph.GA},
       adsurl = {https://ui.adsabs.harvard.edu/abs/2016A&A...594A.116H}
}

@INPROCEEDINGS{NovikovThorne:1973,
       author = {{Novikov}, I.~D. and {Thorne}, K.~S.},
        title = "{Astrophysics of black holes.}",
    booktitle = {Black Holes (Les Astres Occlus)},
         year = 1973,
        month = jan,
        pages = {343-450},
       adsurl = {https://ui.adsabs.harvard.edu/abs/1973blho.conf..343N}
}

@ARTICLE{ShapiroX-rays:1976ApJ,
       author = {{Shapiro}, S.~L. and {Lightman}, A.~P. and {Eardley}, D.~M.},
        title = "{A two-temperature accretion disk model for Cygnus X-1: structure and spectrum.}",
      journal = {\apj},
         year = 1976,
        month = feb,
       volume = {204},
        pages = {187-199},
          doi = {10.1086/154162},
       adsurl = {https://ui.adsabs.harvard.edu/abs/1976ApJ...204..187S}
}

@ARTICLE{Sunyaev_Titarchuk:1980,
       author = {{Sunyaev}, R.~A. and {Titarchuk}, L.~G.},
        title = "{Comptonization of X-Rays in Plasma Clouds - Typical Radiation Spectra}",
      journal = {\aap},
         year = 1980,
        month = jun,
       volume = {86},
        pages = {121},
       adsurl = {https://ui.adsabs.harvard.edu/abs/1980A&A....86..121S}
}

@ARTICLE{Lightman_Zdiarski:1987ApJComptonScattering,
       author = {{Lightman}, Alan P. and {Zdziarski}, Andrzej A.},
        title = "{Pair Production and Compton Scattering in Compact Sources and Comparison to Observations of Active Galactic Nuclei}",
      journal = {\apj},
         year = 1987,
        month = aug,
       volume = {319},
        pages = {643},
          doi = {10.1086/165485},
       adsurl = {https://ui.adsabs.harvard.edu/abs/1987ApJ...319..643L}
}

@article{Dong+Refelction_Disk:2023,
	title = {X-ray {Reflection} from the {Plunging} {Region} of {Black} {Hole} {Accretion} {Disks}},
	url = {http://arxiv.org/abs/2312.09210},
	urldate = {2024-04-19},
	publisher = {arXiv},
      journal = {arXiv e-prints},
	author = {Dong, Jameson and Mastroserio, Guglielmo and Garcıa, Javier A. and Ingram, Adam and Nathan, Edward and Connors, Riley},
	month = dec,
	year = {2023},
	note = {arXiv:2312.09210 [astro-ph]},
}

@article{Goad+BLR:2012,
	title = {The broad emission-line region: the confluence of the outer accretion disc with the inner edge of the dusty torus},
	volume = {426},
	issn = {0035-8711},
	shorttitle = {The broad emission-line region},
	url = {https://ui.adsabs.harvard.edu/abs/2012MNRAS.426.3086G},
	doi = {10.1111/j.1365-2966.2012.21808.x},
	urldate = {2023-02-28},
	journal = {Monthly Notices of the Royal Astronomical Society},
	author = {Goad, M. R. and Korista, K. T. and Ruff, A. J.},
	month = nov,
	year = {2012},
	note = {ADS Bibcode: 2012MNRAS.426.3086G},
	pages = {3086--3111},
}

@article{Saha+:Torus_Xrays:2021,
	title = {Inferring the morphology of {AGN} torus using {X}-ray spectra: {A} reliability study},
	volume = {509},
	issn = {0035-8711, 1365-2966},
	shorttitle = {Inferring the morphology of {AGN} torus using {X}-ray spectra},
	url = {http://arxiv.org/abs/2112.07393},
	doi = {10.1093/mnras/stab3250},
	number = {4},
	urldate = {2022-09-12},
	journal = {Monthly Notices of the Royal Astronomical Society},
	author = {Saha, Tathagata and Markowitz, Alex G. and Buchner, Johannes},
	month = dec,
	year = {2021},
	note = {arXiv:2112.07393 [astro-ph]},
	pages = {5485--5510},
}

@article{Netzer_UniModelreview:2015,
	title = {Revisiting the {Unified} {Model} of {Active} {Galactic} {Nuclei}},
	volume = {53},
	issn = {0066-4146},
	url = {https://ui.adsabs.harvard.edu/abs/2015ARA&A..53..365N},
	doi = {10.1146/annurev-astro-082214-122302},
	urldate = {2024-04-19},
	journal = {Annual Review of Astronomy and Astrophysics},
	author = {Netzer, Hagai},
	month = aug,
	year = {2015},
	note = {ADS Bibcode: 2015ARA\&A..53..365N},
	pages = {365--408},
}

@article{stalevski_dust_2016,
	title = {The dust covering factor in active galactic nuclei},
	volume = {458},
	issn = {0035-8711},
	url = {https://ui.adsabs.harvard.edu/abs/2016MNRAS.458.2288S},
	doi = {10.1093/mnras/stw444},
	urldate = {2024-04-19},
	journal = {Monthly Notices of the Royal Astronomical Society},
	author = {Stalevski, Marko and Ricci, Claudio and Ueda, Yoshihiro and Lira, Paulina and Fritz, Jacopo and Baes, Maarten},
	month = may,
	year = {2016},
	note = {Publisher: OUP
ADS Bibcode: 2016MNRAS.458.2288S},
	pages = {2288--2302},
}

@article{Hao_IR-torus:2005,
doi = {10.1086/431227},
url = {https://dx.doi.org/10.1086/431227},
year = {2005},
month = {may},
publisher = {},
volume = {625},
number = {2},
pages = {L75},
author = {Lei Hao and H. W. W. Spoon and G. C. Sloan and J. A. Marshall and L. Armus and A. G. G. M. Tielens and B. Sargent and I. M. van Bemmel and V. Charmandaris and D. W. Weedman and J. R. Houck},
title = {The Detection of Silicate Emission from Quasars at 10 and 18 Microns},
journal = {The Astrophysical Journal}
}

@ARTICLE{MendozaCastrejon:IRtorus:2015,
       author = {{Mendoza-Castrej{\'o}n}, S. and {Dultzin}, D. and {Krongold}, Y. and {Gonz{\'a}lez}, J.~J. and {Elitzur}, M.},
        title = "{The dust geometric distribution in Seyfert 1 and Seyfert 2 galaxies, isolated and in interaction}",
      journal = {\mnras},
         year = 2015,
        month = mar,
       volume = {447},
       number = {3},
        pages = {2437-2444},
          doi = {10.1093/mnras/stu2566},
       adsurl = {https://ui.adsabs.harvard.edu/abs/2015MNRAS.447.2437M}
}

@ARTICLE{Hatziminaoglou+:MIRtorus:2015ApJ,
       author = {{Hatziminaoglou}, E. and {Hern{\'a}n-Caballero}, A. and {Feltre}, A. and {Pi{\~n}ol Ferrer}, N.},
        title = "{A Complete Census of Silicate Features in the Mid-infrared Spectra of Active Galaxies}",
      journal = {\apj},
         year = 2015,
        month = apr,
       volume = {803},
       number = {2},
          eid = {110},
        pages = {110},
          doi = {10.1088/0004-637X/803/2/110},
archivePrefix = {arXiv},
       eprint = {1502.05823},
 primaryClass = {astro-ph.GA},
       adsurl = {https://ui.adsabs.harvard.edu/abs/2015ApJ...803..110H}
}

@ARTICLE{LopezGonzaga:NGC1068:2014,
       author = {{L{\'o}pez-Gonzaga}, N. and {Jaffe}, W. and {Burtscher}, L. and {Tristram}, K.~R.~W. and {Meisenheimer}, K.},
        title = "{Revealing the large nuclear dust structures in NGC 1068 with MIDI/VLTI}",
      journal = {\aap},
         year = 2014,
        month = may,
       volume = {565},
          eid = {A71},
        pages = {A71},
          doi = {10.1051/0004-6361/201323002},
archivePrefix = {arXiv},
       eprint = {1401.3248},
 primaryClass = {astro-ph.GA},
       adsurl = {https://ui.adsabs.harvard.edu/abs/2014A&A...565A..71L}
}

@ARTICLE{Honig:PolarCone:2013ApJ,
       author = {{H{\"o}nig}, S.~F. and {Kishimoto}, M. and {Tristram}, K.~R.~W. and {Prieto}, M.~A. and {Gandhi}, P. and {Asmus}, D. and {Antonucci}, R. and {Burtscher}, L. and {Duschl}, W.~J. and {Weigelt}, G.},
        title = "{Dust in the Polar Region as a Major Contributor to the Infrared Emission of Active Galactic Nuclei}",
      journal = {\apj},
         year = 2013,
        month = jul,
       volume = {771},
       number = {2},
          eid = {87},
        pages = {87},
          doi = {10.1088/0004-637X/771/2/87},
archivePrefix = {arXiv},
       eprint = {1306.4312},
 primaryClass = {astro-ph.CO},
       adsurl = {https://ui.adsabs.harvard.edu/abs/2013ApJ...771...87H}
}

@article{asmus_PolarCOne:2016,
	title = {The {Subarcsecond} {Mid}-infrared {View} of {Local} {Active} {Galactic} {Nuclei}. {III}. {Polar} {Dust} {Emission}},
	volume = {822},
	issn = {0004-637X},
	url = {https://ui.adsabs.harvard.edu/abs/2016ApJ...822..109A},
	doi = {10.3847/0004-637X/822/2/109},
	urldate = {2022-11-22},
	journal = {The Astrophysical Journal},
	author = {Asmus, D. and Hönig, S. F. and Gandhi, P.},
	month = may,
	year = {2016},
	note = {ADS Bibcode: 2016ApJ...822..109A},
	pages = {109},
}

@article{Stalevski_PolarCone:2017,
	title = {Dissecting the active galactic nucleus in {Circinus} - {I}. {Peculiar} mid-{IR} morphology explained by a dusty hollow cone},
	volume = {472},
	issn = {0035-8711},
	url = {https://ui.adsabs.harvard.edu/abs/2017MNRAS.472.3854S},
	doi = {10.1093/mnras/stx2227},
	urldate = {2023-01-10},
	journal = {Monthly Notices of the Royal Astronomical Society},
	author = {Stalevski, Marko and Asmus, Daniel and Tristram, Konrad R. W.},
	month = dec,
	year = {2017},
	note = {ADS Bibcode: 2017MNRAS.472.3854S},
	pages = {3854--3870},
}

@ARTICLE{DelMoroNustarSurvey:2017ApJ,
       author = {{Del Moro}, A. and {Alexander}, D.~M. and {Aird}, J.~A. and {Bauer}, F.~E. and {Civano}, F. and {Mullaney}, J.~R. and {Ballantyne}, D.~R. and {Brandt}, W.~N. and {Comastri}, A. and {Gandhi}, P. and {Harrison}, F.~A. and {Lansbury}, G.~B. and {Lanz}, L. and {Luo}, B. and {Marchesi}, S. and {Puccetti}, S. and {Ricci}, C. and {Saez}, C. and {Stern}, D. and {Treister}, E. and {Zappacosta}, L.},
        title = "{The NuSTAR Extragalactic Survey: Average Broadband X-Ray Spectral Properties of the NuSTAR-detected AGNs}",
      journal = {\apj},
         year = 2017,
        month = nov,
       volume = {849},
       number = {1},
          eid = {57},
        pages = {57},
          doi = {10.3847/1538-4357/aa9115},
archivePrefix = {arXiv},
       eprint = {1710.01041},
 primaryClass = {astro-ph.HE},
       adsurl = {https://ui.adsabs.harvard.edu/abs/2017ApJ...849...57D}
}

@article{Shu_FEKA_2010,
	title = {{THE} {CORES} {OF} {THE} {Fe} {Kα} {LINES} {IN} {ACTIVE} {GALACTIC} {NUCLEI}: {AN} {EXTENDED} {CHANDRA} {HIGH} {ENERGY} {GRATING} {SAMPLE}},
	volume = {187},
	issn = {0067-0049},
	shorttitle = {{THE} {CORES} {OF} {THE} {Fe} {Kα} {LINES} {IN} {ACTIVE} {GALACTIC} {NUCLEI}},
	url = {https://dx.doi.org/10.1088/0067-0049/187/2/581},
	doi = {10.1088/0067-0049/187/2/581},
	language = {en},
	number = {2},
	urldate = {2023-09-15},
	journal = {The Astrophysical Journal Supplement Series},
	author = {Shu, X. W. and Yaqoob, T. and Wang, J. X.},
	month = mar,
	year = {2010},
	note = {Publisher: The American Astronomical Society},
	pages = {581},
}

@ARTICLE{Fabian_ComptonHump:1991MNRAS,
       author = {{George}, I.~M. and {Fabian}, A.~C.},
        title = "{X-ray reflection from cold matter in Active Galactic Nuclei and X-ray binaries.}",
      journal = {\mnras},
         year = 1991,
        month = mar,
       volume = {249},
        pages = {352},
          doi = {10.1093/mnras/249.2.352},
       adsurl = {https://ui.adsabs.harvard.edu/abs/1991MNRAS.249..352G}
}

@misc{Haidar+JWSTcone:2024ARXIV,
	title = {Dust beyond the torus: {Revealing} the mid-infrared heart of local {Seyfert} {ESO} 428-{G14} with {JWST}/{MIRI}},
	shorttitle = {Dust beyond the torus},
	url = {http://arxiv.org/abs/2404.16100},
	language = {en},
	urldate = {2024-04-26},
	publisher = {arXiv},
	author = {Haidar, Houda and Rosario, David J. and Alonso-Herrero, Almudena and Pereira-Santaella, Miguel and García-Bernete, Ismael and Campbell, Stephanie and Hönig, Sebastian F. and Almeida, Cristina Ramos and Hicks, Erin and Delaney, Daniel and Davies, Richard and Ricci, Claudio and Harrison, Chris M. and Leist, Mason and Lopez-Rodriguez, Enrique and Garcia-Burillo, Santiago and Zhang, Lulu and Packham, Chris and Gandhi, Poshak and Audibert, Anelise and Bellocchi, Enrica and Boorman, Peter and Bunker, Andrew and Combes, Françoise and Santos, Tanio Diaz and Donnan, Fergus R. and Martin, Omaira Gonzalez and Muñoz, Laura Hermosa and Charidis, Matthaios and Labiano, Alvaro and Levenson, Nancy A. and May, Daniel and Rigopoulou, Dimitra and Ardila, Alberto Rodriguez and Shimizu, T. Taro and Stalevski, Marko and Ward, Martin},
	month = apr,
	year = {2024},
	note = {arXiv:2404.16100 [astro-ph]},
}

@ARTICLE{Kormendy_Review_SMBH:2013,
       author = {{Kormendy}, John and {Ho}, Luis C.},
        title = "{Coevolution (Or Not) of Supermassive Black Holes and Host Galaxies}",
      journal = {\araa},
         year = 2013,
        month = aug,
       volume = {51},
       number = {1},
        pages = {511-653},
          doi = {10.1146/annurev-astro-082708-101811},
archivePrefix = {arXiv},
       eprint = {1304.7762},
 primaryClass = {astro-ph.CO},
       adsurl = {https://ui.adsabs.harvard.edu/abs/2013ARA&A..51..511K}
}

@ARTICLE{Marconi_BH_demographics:2004,
       author = {{Marconi}, A. and {Risaliti}, G. and {Gilli}, R. and {Hunt}, L.~K. and {Maiolino}, R. and {Salvati}, M.},
        title = "{Local supermassive black holes, relics of active galactic nuclei and the X-ray background}",
      journal = {\mnras},
         year = 2004,
        month = jun,
       volume = {351},
       number = {1},
        pages = {169-185},
          doi = {10.1111/j.1365-2966.2004.07765.x},
archivePrefix = {arXiv},
       eprint = {astro-ph/0311619},
 primaryClass = {astro-ph},
       adsurl = {https://ui.adsabs.harvard.edu/abs/2004MNRAS.351..169M}
}

@ARTICLE{Reynolds_reflection_spin:2008ApJ,
       author = {{Reynolds}, Christopher S. and {Fabian}, Andrew C.},
        title = "{Broad Iron-K{\ensuremath{\alpha}} Emission Lines as a Diagnostic of Black Hole Spin}",
      journal = {\apj},
         year = 2008,
        month = mar,
       volume = {675},
       number = {2},
        pages = {1048-1056},
          doi = {10.1086/527344},
archivePrefix = {arXiv},
       eprint = {0711.4158},
 primaryClass = {astro-ph},
       adsurl = {https://ui.adsabs.harvard.edu/abs/2008ApJ...675.1048R}
}

@ARTICLE{Diaz_reflection_LLAGN:2023AA,
       author = {{Diaz}, Y. and {Hern{\`a}ndez-Garc{\'\i}a}, L. and {Ar{\'e}valo}, P. and {L{\'o}pez-Navas}, E. and {Ricci}, C. and {Koss}, M. and {Gonzalez-Martin}, O. and {Balokovi{\'c}}, M. and {Osorio-Clavijo}, N. and {Garc{\'\i}a}, J.~A. and {Malizia}, A.},
        title = "{Constraining the X-ray reflection in low accretion-rate active galactic nuclei using XMM-Newton, NuSTAR, and Swift}",
      journal = {\aap},
         year = 2023,
        month = jan,
       volume = {669},
          eid = {A114},
        pages = {A114},
          doi = {10.1051/0004-6361/202244678},
archivePrefix = {arXiv},
       eprint = {2210.15376},
 primaryClass = {astro-ph.HE},
       adsurl = {https://ui.adsabs.harvard.edu/abs/2023A&A...669A.114D}
}

@ARTICLE{BrightmanNandraXMMsurvey:2011MNRAS,
       author = {{Brightman}, Murray and {Nandra}, Kirpal},
        title = "{An XMM-Newton spectral survey of 12 {\ensuremath{\mu}}m selected galaxies - I. X-ray data}",
      journal = {\mnras},
         year = 2011,
        month = may,
       volume = {413},
       number = {2},
        pages = {1206-1235},
          doi = {10.1111/j.1365-2966.2011.18207.x},
archivePrefix = {arXiv},
       eprint = {1012.3345},
 primaryClass = {astro-ph.HE},
       adsurl = {https://ui.adsabs.harvard.edu/abs/2011MNRAS.413.1206B}
}

@ARTICLE{Salpeter_earlySMBH:1964ApJ,
       author = {{Salpeter}, E.~E.},
        title = "{Accretion of Interstellar Matter by Massive Objects.}",
      journal = {\apj},
         year = 1964,
        month = aug,
       volume = {140},
        pages = {796-800},
          doi = {10.1086/147973},
       adsurl = {https://ui.adsabs.harvard.edu/abs/1964ApJ...140..796S}
}

@ARTICLE{Lynden-Bell_earlyQSO:1969Natur,
       author = {{Lynden-Bell}, D.},
        title = "{Galactic Nuclei as Collapsed Old Quasars}",
      journal = {\nat},
         year = 1969,
        month = aug,
       volume = {223},
       number = {5207},
        pages = {690-694},
          doi = {10.1038/223690a0},
       adsurl = {https://ui.adsabs.harvard.edu/abs/1969Natur.223..690L}
}

@ARTICLE{CavalierePadovani_AGNvsNormal:1989ApJ,
       author = {{Cavaliere}, A. and {Padovani}, P.},
        title = "{The Connection between Active and Normal Galaxies}",
      journal = {\apjl},
         year = 1989,
        month = may,
       volume = {340},
        pages = {L5},
          doi = {10.1086/185425},
       adsurl = {https://ui.adsabs.harvard.edu/abs/1989ApJ...340L...5C}
}

@article{GravityNGC1068:2020,
	title = {An image of the dust sublimation region in the nucleus of {NGC} 1068},
	volume = {634},
	issn = {0004-6361, 1432-0746},
	url = {http://arxiv.org/abs/1912.01361},
	doi = {10.1051/0004-6361/201936255},
	language = {en},
	urldate = {2024-05-23},
	journal = {Astronomy \& Astrophysics},
	author = {{GRAVITY Collaboration} and Pfuhl, O. and Davies, R. and Dexter, J. and Netzer, H. and Hoenig, S. and Lutz, D. and Schartmann, M. and Sturm, E. and Amorim, A. and Brandner, W. and Clenet, Y. and de Zeeuw, P. T. and Eckart, A. and Eisenhauer, F. and Schreiber, N. M. Foerster and Gao, F. and Garcia, P. J. V. and Genzel, R. and Gillessen, S. and Gratadour, D. and Kishimoto, M. and Lacour, S. and Millour, F. and Ott, T. and Paumard, T. and Perraut, K. and Perrin, G. and Peterson, B. M. and Petrucci, P. O. and Prieto, M. A. and Rouan, D. and Shangguan, J. and Shimizu, T. and Sternberg, A. and Straub, O. and Straubmeier, C. and Tacconi, L. J. and Tristram, K. R. W. and Vermot, P. and Waisberg, I. and Widmann, F. and Woillez, J.},
	month = feb,
	year = {2020},
	note = {arXiv:1912.01361 [astro-ph]},
	pages = {A1}
}

@ARTICLE{GATOSII:2021A&A,
       author = {{Alonso-Herrero}, A. and {Garc{\'\i}a-Burillo}, S. and {H{\"o}nig}, S.~F. and {Garc{\'\i}a-Bernete}, I. and {Ramos Almeida}, C. and {Gonz{\'a}lez-Mart{\'\i}n}, O. and {L{\'o}pez-Rodr{\'\i}guez}, E. and {Boorman}, P.~G. and {Bunker}, A.~J. and {Burtscher}, L. and {Combes}, F. and {Davies}, R. and {D{\'\i}az-Santos}, T. and {Gandhi}, P. and {Garc{\'\i}a-Lorenzo}, B. and {Hicks}, E.~K.~S. and {Hunt}, L.~K. and {Ichikawa}, K. and {Imanishi}, M. and {Izumi}, T. and {Labiano}, A. and {Levenson}, N.~A. and {Packham}, C. and {Pereira-Santaella}, M. and {Ricci}, C. and {Rigopoulou}, D. and {Roche}, P. and {Rosario}, D.~J. and {Rouan}, D. and {Shimizu}, T. and {Stalevski}, M. and {Wada}, K. and {Williamson}, D.},
        title = "{The Galaxy Activity, Torus, and Outflow Survey (GATOS). II. Torus and polar dust emission in nearby Seyfert galaxies}",
      journal = {\aap},
         year = 2021,
        month = aug,
       volume = {652},
          eid = {A99},
        pages = {A99},
          doi = {10.1051/0004-6361/202141219},
archivePrefix = {arXiv},
       eprint = {2107.00244},
 primaryClass = {astro-ph.GA},
       adsurl = {https://ui.adsabs.harvard.edu/abs/2021A&A...652A..99A}
}

@INPROCEEDINGS{Rowan-Robinson:1998ASPC,
       author = {{Rowan-Robinson}, M.},
        title = "{Infrared Luminous Galaxies and AGN}",
    booktitle = {Science With The NGST},
         year = 1998,
       editor = {{Smith}, Eric P. and {Koratkar}, Anuradha},
       series = {Astronomical Society of the Pacific Conference Series},
       volume = {133},
        month = jan,
        pages = {119},
       adsurl = {https://ui.adsabs.harvard.edu/abs/1998ASPC..133..119R}
}

@ARTICLE{Rowan-RobinsonIRmodels:1995MNRAS,
       author = {{Rowan-Robinson}, Michael},
        title = "{A new model for the infrared emission of quasars}",
      journal = {\mnras},
         year = 1995,
        month = feb,
       volume = {272},
       number = {4},
        pages = {737-748},
          doi = {10.1093/mnras/272.4.737},
       adsurl = {https://ui.adsabs.harvard.edu/abs/1995MNRAS.272..737R}
}

@article{Alexander_2013CXB+Reflection:2013,
doi = {10.1088/0004-637X/773/2/125},
url = {https://dx.doi.org/10.1088/0004-637X/773/2/125},
year = {2013},
month = {aug},
publisher = {The American Astronomical Society},
volume = {773},
number = {2},
pages = {125},
author = {D. M. Alexander and D. Stern and A. Del Moro and G. B. Lansbury and R. J. Assef and J. Aird and M. Ajello and D. R. Ballantyne and F. E. Bauer and S. E. Boggs and W. N. Brandt and F. E. Christensen and F. Civano and A. Comastri and W. W. Craig and M. Elvis and B. W. Grefenstette and C. J. Hailey and F. A. Harrison and R. C. Hickox and B. Luo and K. K. Madsen and J. R. Mullaney and M. Perri and S. Puccetti and C. Saez and E. Treister and C. M. Urry and W. W. Zhang and C. R. Bridge and P. R. M. Eisenhardt and A. H. Gonzalez and S. H. Miller and C. W. Tsai},
title = {THE NuSTAR EXTRAGALACTIC SURVEY: A FIRST SENSITIVE LOOK AT THE HIGH-ENERGY COSMIC X-RAY BACKGROUND POPULATION},
journal = {The Astrophysical Journal}
}

@ARTICLE{Lamer_Uttley_RXTEreflection:2000MNRAS,
       author = {{Lamer}, G. and {Uttley}, P. and {McHardy}, I.~M.},
        title = "{RXTE observations of the Seyfert galaxy NGC 5506: evidence for reflection from disc and torus}",
      journal = {\mnras},
         year = 2000,
        month = dec,
       volume = {319},
       number = {3},
        pages = {949-955},
          doi = {10.1046/j.1365-8711.2000.03921.x},
archivePrefix = {arXiv},
       eprint = {astro-ph/0008219},
 primaryClass = {astro-ph},
       adsurl = {https://ui.adsabs.harvard.edu/abs/2000MNRAS.319..949L}
}

@ARTICLE{George_Fabian_reflection:1991MNRAS,
       author = {{George}, I.~M. and {Fabian}, A.~C.},
        title = "{X-ray reflection from cold matter in Active Galactic Nuclei and X-ray binaries.}",
      journal = {\mnras},
         year = 1991,
        month = mar,
       volume = {249},
        pages = {352},
          doi = {10.1093/mnras/249.2.352},
       adsurl = {https://ui.adsabs.harvard.edu/abs/1991MNRAS.249..352G}
}

@ARTICLE{Magdziarz_Zdiarski_reflection:1995MNRAS,
       author = {{Magdziarz}, Pawel and {Zdziarski}, Andrzej A.},
        title = "{Angle-dependent Compton reflection of X-rays and gamma-rays}",
      journal = {\mnras},
         year = 1995,
        month = apr,
       volume = {273},
       number = {3},
        pages = {837-848},
          doi = {10.1093/mnras/273.3.837},
       adsurl = {https://ui.adsabs.harvard.edu/abs/1995MNRAS.273..837M}
}

@ARTICLE{Yaqoob_Serlemitos_FEKA3C273_2000ApJ,
       author = {{Yaqoob}, Tahir and {Serlemitsos}, Peter},
        title = "{A Broad Fe K{\ensuremath{\alpha}} Emission Line in the X-Ray Spectrum of the Quasar 3C 273}",
      journal = {\apjl},
         year = 2000,
        month = dec,
       volume = {544},
       number = {2},
        pages = {L95-L99},
          doi = {10.1086/317318},
archivePrefix = {arXiv},
       eprint = {astro-ph/0009435},
 primaryClass = {astro-ph},
       adsurl = {https://ui.adsabs.harvard.edu/abs/2000ApJ...544L..95Y}
}

@ARTICLE{Nandra_GeorgeASCA_FEKA+:1997ApJ,
       author = {{Nandra}, K. and {George}, I.~M. and {Mushotzky}, R.~F. and {Turner}, T.~J. and {Yaqoob}, T.},
        title = "{ASCA Observations of Seyfert 1 Galaxies. II. Relativistic Iron K{\ensuremath{\alpha}} Emission}",
      journal = {\apj},
         year = 1997,
        month = mar,
       volume = {477},
       number = {2},
        pages = {602-622},
          doi = {10.1086/303721},
archivePrefix = {arXiv},
       eprint = {astro-ph/9606169},
 primaryClass = {astro-ph},
       adsurl = {https://ui.adsabs.harvard.edu/abs/1997ApJ...477..602N}
}
\bibliographystyle{aasjournal}

\appendix

\section{The addition of the polar cone component}\label{app:cone_vs_torus}

\begin{figure*}[ht!]
    \centering
    \includegraphics[width=0.9\textwidth]{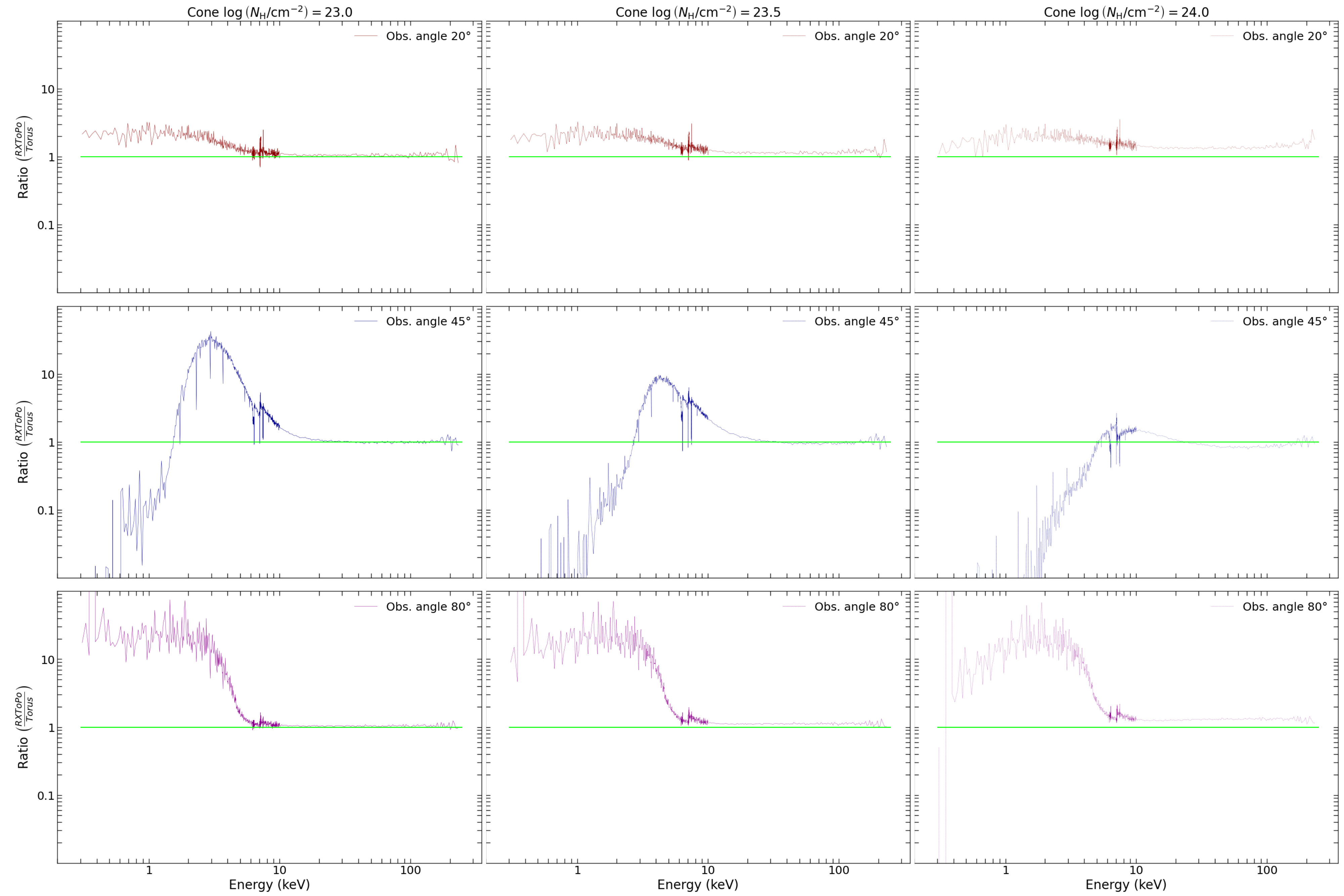}
    \caption{A comparison between the retrieved spectra. The spectra are continuum subtracted and then the ratio has been calculated for two configuration. First, a simple torus with equatorial column density $N_{\rm H}= 10^{24}\,{\rm cm^{-2}}$ and covering factor 0.6. Second, the \rxtopo{} with the same torus and three different polar \textbf{cone column densities ($\log\left(N_{\rm H}/{\rm cm^{-2}}\right)= 23.0-23.5-24.0$)}. The three observing angles are selected so that: At $20^{\circ}$ no medium is intercepted, at $45^{\circ}$ the cone is intercepted and $80^{\circ}$ the torus is intercepted.}
    \label{fig:torus_vs_cone}
\end{figure*}

In this work, we add the hollow polar cone component perpendicular to the torus plane.
The addition of such a medium affects the reflection, and the emerging spectrum can vary from a simple toroidal configuration.
Hence, we make some comparisons between a simple torus spectrum and the \rxtopo{}.
In both cases, we collect only reprocessed photons ("RPRC" see Sec.\,\ref{sec:different_components}) and we present their ratio in Fig.\,\ref{fig:torus_vs_cone}.
The torus properties are fixed in all cases ($N_{\rm H}= 10^{24}\,{\rm cm^{-2}}$ and covering factor 0.6).
For the cone, we use three different column density options, and the observing angle varies so that i) no material is intercepted ($20^{\circ}$), ii) the cone is intercepted ($45^{\circ}$) and iii) the torus is intercepted ($80^{\circ}$).
The results suggest that in low observing angles, where no material is in the line of sight, the two spectra are similar, with a slight increase in the reflection due to the cone.
In medium angles, there is stronger absorption in low energies and a bump in mid energies ($\rm 1\,keV < energy < 10\,keV$) because of the enhanced reflection in these energies.
In high observing angles, the torus absorption dominates.
This generates an increase in the lower energies, which consists of photons that otherwise would have been lost and now are reflected towards the collector.
With this test, we illustrate that the polar dusty cone is indeed a useful component that can affect the observed emerged spectrum.

\section{Line-of-sight column density}\label{app:reflexino}

The best way to calculate the column density that the photons intercept along the line of sight is to use the \texttt{REFLEXINO} tool.
In this demo, we present how you can use this tool, based on the fit found in Sec.\,\ref{sec:fitNGC424} for the \rxtopo{} model.
We need to develop a parameter file (for example, \texttt{my\_model.par}) with the proper commands and parameters (for more details, see the manual of \Reflex{}\footnote{\url{https://www.astro.unige.ch/reflex/reflex-software}}).
Here is an example that can be used to calculate the line-of-sight column density for the \rxtopo{} best fit:
\begin{verbatim}
    VERBOSE 0
    NPHOTS 100000
    LENGTH Centimeter
    EMSPEC REFLEXINO
    EMGEOM SPHERE 0.0 0.0 5.01534e12 3.0092e12 0.0 0.0 90
    OBJECT WORLD 3e30
    #OBJECT TORUS EXTERNAL
    MATTER lpgs
    DUST 1.0
    TEMPERATURE 0
    H2FRACTION 0.3
    METALLICITY 1
    DENSITY 2503429.94 #change it accordingly= nh*(2x INNER_RADIUS)
    OBJECT TORUS TOR 0.0 0.0 0.0 9.1388e+17 6.0316e+17
    #OBJECT HOLLOW CONE
    MATTER lpgs
    DUST 1.0
    TEMPERATURE 0
    H2FRACTION 0.3
    METALLICITY 1
    DENSITY 995.366 #for NH 1e23.19
    OBJECT HCONE HCONE 0.0 0.0 0.0 3.1072e+17 1.2343e+20 3.4148e+17 1.3565+20 2.4015e+17 9.5396e+19
    PERCENT ON
    HISTOGRAM BINNING 20.0 25.5 0.001 LOG
    HISTOGRAM NEW NH NHdeg79.txt
    HISTOGRAM DIR_Z < 0.20791169081775945 #78deg
    HISTOGRAM DIR_Z > 0.19080899537654492 #79deg
\end{verbatim}

The output of this simulation is a text file, in this example \texttt{NHdeg79.txt}, which consists of three columns. The first two columns are the left and right bins of the column density, and the third column gives the number of photons that have been intercepted in that bin.
Normally, the third column should be full of zeros except the bin in which the line-of-sight column density is calculated.
In some cases a few bins before and after can be populated with a few photons, but this is not a problem.
We consider the bin with the maximum number of photons to be the correct one.
In this case, the line-of-sight column density is $N_{\rm H}^{\rm los} = 2.89 \times 10^{24}\,{\rm cm^{-2}}$.

In the case of the \rxagn{} model, we need to include the BLR complex and then follow the same procedure.
\begin{verbatim}
DENSITY 3.2e+09
OBJECT ANNULUS FLAR_DSK 0.0 0.0 0.0 5.6458e+15 5.0119e+15 1.0000e+13
DENSITY 2.5e+09
OBJECT ANNULUS FLAR_DSK 0.0 0.0 0.0 6.3599e+15 5.6458e+15 1.3705e+13
DENSITY 1.9e+09
OBJECT ANNULUS FLAR_DSK 0.0 0.0 0.0 7.1643e+15 6.3599e+15 1.8783e+13
DENSITY 1.5e+09
OBJECT ANNULUS FLAR_DSK 0.0 0.0 0.0 8.0704e+15 7.1643e+15 2.5742e+13
DENSITY 1.1e+09
OBJECT ANNULUS FLAR_DSK 0.0 0.0 0.0 9.0912e+15 8.0705e+15 3.5279e+13
DENSITY 8.9e+08
OBJECT ANNULUS FLAR_DSK 0.0 0.0 0.0 1.0241e+16 9.0913e+15 4.8350e+13
DENSITY 6.9e+08
OBJECT ANNULUS FLAR_DSK 0.0 0.0 0.0 1.1536e+16 1.0241e+16 6.6263e+13
DENSITY 5.3e+08
OBJECT ANNULUS FLAR_DSK 0.0 0.0 0.0 1.2996e+16 1.1536e+16 9.0814e+13
DENSITY 4.1e+08
OBJECT ANNULUS FLAR_DSK 0.0 0.0 0.0 1.4639e+16 1.2996e+16 1.2446e+14
DENSITY 3.2e+08
OBJECT ANNULUS FLAR_DSK 0.0 0.0 0.0 1.6491e+16 1.4639e+16 1.7057e+14
DENSITY 2.5e+08
OBJECT ANNULUS FLAR_DSK 0.0 0.0 0.0 1.8577e+16 1.6491e+16 2.3377e+14
DENSITY 1.9e+08
OBJECT ANNULUS FLAR_DSK 0.0 0.0 0.0 2.0926e+16 1.8577e+16 3.2038e+14
DENSITY 1.5e+08
OBJECT ANNULUS FLAR_DSK 0.0 0.0 0.0 2.3573e+16 2.0926e+16 4.3908e+14
DENSITY 1.2e+08
OBJECT ANNULUS FLAR_DSK 0.0 0.0 0.0 2.6554e+16 2.3573e+16 6.0176e+14
DENSITY 9.0e+07
OBJECT ANNULUS FLAR_DSK 0.0 0.0 0.0 2.9913e+16 2.6555e+16 8.2472e+14
DENSITY 7.0e+07
OBJECT ANNULUS FLAR_DSK 0.0 0.0 0.0 3.3697e+16 2.9913e+16 1.1303e+15
DENSITY 5.4e+07
OBJECT ANNULUS FLAR_DSK 0.0 0.0 0.0 3.7959e+16 3.3697e+16 1.5490e+15
DENSITY 4.2e+07
OBJECT ANNULUS FLAR_DSK 0.0 0.0 0.0 4.2760e+16 3.7959e+16 2.1230e+15
DENSITY 3.3e+07
OBJECT ANNULUS FLAR_DSK 0.0 0.0 0.0 4.8168e+16 4.2760e+16 2.9095e+15
DENSITY 2.5e+07
OBJECT ANNULUS FLAR_DSK 0.0 0.0 0.0 5.4260e+16 4.8168e+16 3.9875e+15
DENSITY 2.0e+07
OBJECT ANNULUS FLAR_DSK 0.0 0.0 0.0 6.1123e+16 5.4261e+16 5.4649e+15
DENSITY 1.5e+07
OBJECT ANNULUS FLAR_DSK 0.0 0.0 0.0 6.8854e+16 6.1124e+16 7.4896e+15
DENSITY 1.2e+07
OBJECT ANNULUS FLAR_DSK 0.0 0.0 0.0 7.7563e+16 6.8855e+16 1.0265e+16
DENSITY 9.2e+06
OBJECT ANNULUS FLAR_DSK 0.0 0.0 0.0 8.7373e+16 7.7564e+16 1.4067e+16
DENSITY 7.1e+06
OBJECT ANNULUS FLAR_DSK 0.0 0.0 0.0 9.8425e+16 8.7374e+16 1.9280e+16
DENSITY 5.5e+06
OBJECT ANNULUS FLAR_DSK 0.0 0.0 0.0 1.1087e+17 9.8426e+16 2.6423e+16
DENSITY 4.3e+06
OBJECT ANNULUS FLAR_DSK 0.0 0.0 0.0 1.2490e+17 1.1087e+17 3.6212e+16
DENSITY 3.3e+06
OBJECT ANNULUS FLAR_DSK 0.0 0.0 0.0 1.4069e+17 1.2490e+17 4.9629e+16
DENSITY 2.6e+06
OBJECT ANNULUS FLAR_DSK 0.0 0.0 0.0 1.5849e+17 1.4070e+17 6.8016e+16
DENSITY 2.0e+06
OBJECT ANNULUS FLAR_DSK 0.0 0.0 0.0 3.1072e+17 1.5849e+17 9.3216e+16
\end{verbatim}

\end{document}